\documentclass[conference]{IEEEtran}
\usepackage[T1]{fontenc}

\usepackage[utf8]{inputenc}
\usepackage[parfill]{parskip}
\usepackage{float}

\usepackage[hidelinks]{hyperref}

\usepackage{graphicx}

\usepackage{amsmath} 
\usepackage[ruled,vlined,noend]{algorithm2e}
\usepackage{booktabs} 

\begin{document}

\hypersetup{pageanchor=false}

\title{A Cloud Resources Portfolio Optimization Business Model - From Theory to Practice}
\author{\IEEEauthorblockN{Valentin Haag}
\IEEEauthorblockA{\textit{Faculty of Computer Science} \\
\textit{University of Vienna}\\
Vienna, Austria\\
valentin.haag94@gmail.com}
\and
\IEEEauthorblockN{Maximilian Kiessler}
\IEEEauthorblockA{\textit{Faculty of Computer Science} \\
\textit{University of Vienna}\\
Vienna, Austria \\
m.kiessler@hotmail.com}
\and
\IEEEauthorblockN{Benedikt Pittl}
\IEEEauthorblockA{\textit{Faculty of Computer Science} \\
\textit{University of Vienna}\\
Vienna, Austria \\
benedikt.pittl@univie.ac.at}
\and
\IEEEauthorblockN{Erich Schikuta}
\IEEEauthorblockA{\textit{Faculty of Computer Science} \\
\textit{University of Vienna}\\
Vienna, Austria \\
erich.schikuta@univie.ac.at}
}

\hypersetup{pageanchor=true}

%
%
%
%
%
%

\maketitle

\begin{abstract}
Cloud resources have become increasingly important, with many businesses using cloud solutions to supplement or outright replace their existing IT infrastructure. However, as there is a plethora of providers with varying products, services, and markets, it has become increasingly more challenging to keep track of the best solutions for each application. Cloud service intermediaries aim to alleviate this problem by offering services that help users meet their requirements. 

This paper aims to lay the groundwork for developing a cloud portfolio management platform and its business model, defined via a business model canvas. Furthermore, a prototype of a platform is developed offering a cloud portfolio optimization service, using two algorithms developed in previous research to create suitable and well-utilized allocations for a customer's applications. 

\end{abstract}


\section{Introduction}
Over the past years, the cloud resource market has been one of the fastest-growing IT segments. The biggest providers, Amazon Web Services (AWS), Microsoft Azure, and Google Cloud, have seen annual sales increases of over 20\% for several years. Just AWS itself reported a revenue of 80 Billion US dollars for the year 2022, and the entire cloud market achieved revenue of 545.8 Billion dollars worldwide with 22\% growth compared to 2021~\cite{cloud_2022_aws}. This large and expanding market has resulted not only in a growing cloud service provider (from here on, often referred to as CSPs) market but also in several methods of delivery of cloud resources to the customer. 

As mentioned by Pittl~\cite{pittl2019cost}, one big challenge facing both industry and academia is finding a cost-effective solution when buying cloud capacities. Pittls' study concludes that almost all observed resource requests were oversized and offered significant cost reduction potential, which lies not only in reducing the amount of resources bought but also in their composition. Depending on the planning period of the operations to be performed, a different mix of procurement from different market spaces is optimal. The author suggests tackling this problem via a cloud resource trading intermediary.

Offering a service that aids businesses in managing their cloud portfolio and proposing efficient allocations to run their applications, even across various providers, is of great interest in the market. While most larger CSPs offer some functionalities and services that claim to help prevent over-provisioning, like AWS Lambda and Fargate, it is ultimately not in the provider's best interest to reduce the costs for the customer. For the same reason, offering cross-platform support should not be expected of them either.

One way to better leverage the opportunities of the cloud market and make the market more accessible is the use of cloud intermediaries between the CSPs and the customer. Given that, this paper aims to tackle two main research questions:

\begin{itemize}
    \item \emph{Under what kind of business model could such an intermediary operate?} To answer this question, a detailed proposal and description of a viable business model for a cloud resource intermediary will be presented. Intermediaries in a similar fashion have been put forward, but as far as the authors' best knowledge, no conclusive business model exists on how these intermediaries could operate.

    \item \emph{How could such a platform be implemented?} The other goal of this work is the design and implementation of a cloud resource intermediary.
\end{itemize}

Thus, the paper is structured as follows: The next \autoref{state_of_the_art} gives the reader a literature survey on the target research area.
Section~\ref{cloud_portfolio_management} focuses on cloud portfolio optimization theory, and two respective algorithms, a genetic and a greedy one, are described and their performance evaluated.
In \autoref{buiness_model}, we propose our proposal of a business model for a cloud portfolio management platform, consisting of the nine building blocks defined by the business model canvas framework of Osterwalder~\cite{osterwalderbusinessmodel}.
Based on the business model description, we present the implementation of our respective prototype in \autoref{cloud_portfolio_manager}.
The following \autoref{ux_evaluation} presents our application's user experience (UX) evaluation.
Finally, \autoref{conclusion_future_work} summarises the paper's findings, including what future work to expand upon the topic.

\section{State of the Art}\label{state_of_the_art}
The cloud market currently consists of many cloud service providers, each featuring various ways and markets for customers to access their products. There is no single space from which all suppliers can be accessed. The market is an oligopoly, dominated by the most prominent three players, Amazon Cloud Services (AWS), Microsoft Azure, and Google Cloud, according to~\cite{cloudproviders2023}. In the course of this work, we focus on AWS, the largest cloud service provider today, which offers three different resource markets with varying pricing models, which are as follows:
\begin{itemize}
\item \emph{On-demand marketplace}: For maximum flexibility and easy scalability with no advance notice, instances bought on this market are handled in the classic pay-per-use model, where
customers can purchase computing power on-demand and by time period.
\item \emph{Saving plans marketplace}: This model offers significant price reductions (e.g., up to 72\% cheaper) for the customer in exchange for a long-term commitment (e.g., 1- or 3-year plans) to a certain amount of usage.
\item \emph{Spot marketplace}: This market allows customers to benefit from the varying loads that AWS experiences on its services. Amazon EC2 offers instances of their currently unused capacity with even greater discounts of up to 90\% reduced prices compared to the on-demand market.
\end{itemize}

The definition of a business model has not only been one of the earliest focuses of research, but also one of the most hotly debated. Almost every paper concerning this topic has defined its own take on this task, and to this day, there is no general agreement on a universally accepted definition. We follow the definition of~\cite{osterwalderbusinessmodel}, where a business model describes the rationale of how an organization creates, delivers, and captures value.

Similar to the definition of the term, there is also a wide array of frameworks describing the elements a business model comprises. The paper will focus on the Business Model Canvas from Osterwalder and Pigneur~\cite{osterwalderbusinessmodel}, one of the most established and widely used frameworks today. This framework proposes to describe any business model with nine building blocks, which can also be grouped into four areas of a business: customers, offer, infrastructure, and financial viability. They showcase how a company intends to make money and are as follows~\cite{Conceptualising_Business_Models,osterwalderbusinessmodel}:
\begin{itemize}
\item The \emph{Customer Segment} block describes the groups a company tries to offer its products to.
\item The \emph{Value Proposition} describes the reason why a customer chooses to work with one business over others
\item \emph{Channels} describe the way a company reaches its customer segment to make its proposition of value.
\item The different kinds of relationships companies can have with their customers are described in the block \emph{Customer Relationships}.
\item The block \emph{Revenue Streams} deals with the income a company receives from its customers.
\item The \emph{Key Resources} block describes the resources that allow the company to operate its business and earn revenues.
\item The block \emph{Key Activities} defines the activities a company must perform to be successful. They are necessary to create value and earn revenue, and they can differ widely depending on the company.
\item \emph{Key partnerships} are the suppliers and partners a business model includes to make it work.
\item The last block \emph{Cost Structure} of the Canvas deals with all operating costs of the business model.
\end{itemize}


As many cloud providers are on the market nowadays, the question arises: What makes such a company and its business model successful?
An analysis of the business model characteristics revealed three different types of business models, with differing value creation, proposition, and delivery that all providers could be divided into~\cite{successful_bm_cloud_provider}:

\begin{itemize}
    \item \emph{Newcomer:} This type describes newcomers to the market that adapt already existing cloud strategies or form co-operations with already established cloud companies. Their value proposition focuses much on individual customization, often resulting in higher initial costs. The newcomer's main customer segment is niche markets, which they reach through traditional channels, such as print media and personal contact. A one-time charge generates revenue, which supplementary services or partner revenue models later generate. This cloud provider has to deal with a crowded market and stand out by having a well-developed market entry strategy and a defined target market. On the other hand, they have the advantage of working flexibly and developing a specialized cloud service instead of focusing on commodity services.

    \item \emph{Experienced player:} Companies of this type mainly work with their existing know-how, offering a plethora of software and consultancy services. The services offered are standardized and provided via a public cloud with high-security standards, resulting in excellent scalability and possibilities for time and cost reductions. The experienced players target both the mass and individual markets and use support systems and online communities to reach their customer base. Subscription-based services usually account for the biggest source of revenue. While profiting from an economy of scale and a good understanding of their technology, they must be careful that good CRM, branding, and marketing strategies compensate for less direct customer contact and lower trust levels.

    \item \emph{Specialised provider:} Providers of this type stand out by expanding the usual cloud services on a vertical level, adding services such as data processing, administration, marketplace, and migration services. Their target customer segment is branch-specific and may include the public sector. Firms following this business model often implement a usage-based revenue model. The specialized providers have the advantage of a high-quality and innovative model that, in combination with good security and customer orientation, results in high levels of trust and loyalty by their customers. While not easily imitated, the model does not support high scalability and needs constant innovation to satisfy the customers' ever-developing needs.
\end{itemize}

Actual cloud broker implementations on the market are still sparse, while the high number of cloud service offerings makes it hard to find the right service as a customer. Brokerage can be several services the intermediary offers the cloud customer, such as decision support or enhancing the delivery of services. The benefits for the customer are lower costs and the ability to seamlessly switch between different cloud providers~\cite{cloud_broker_survey}.

To gain a deeper understanding of the cloud intermediary market, Elhabbash et al.~\cite{cloud_broker_survey} presents a systematic survey of cloud brokerage literature, looking into the motivation for designing a cloud broker and its functionality.

While a plethora of research has been conducted in the field of cloud intermediaries, the number of implementations on the market is limited. This situation, in conjunction with the fact that the services provided can vary wildly, has resulted in a lack of research concerning the business models of cloud brokers so far. Nevertheless, the work of Filiopoulou at al.~\cite{cloud_brokerage_business_model} gives an overview of benefits, common pricing models, and an evaluation of cloud brokers. It is concluded that brokers assist companies in developing themselves and creating a more competitive environment for providers while earning revenues themselves.

\section{Portfolio Optimization Model}\label{cloud_portfolio_management}
In this section, we will present a synopsis of the findings of our previous research work, which encompasses related approaches, a formulation of the problem at hand, a short description of two portfolio optimization algorithms developed, and finally, an evaluation of their performance~\cite{kiessler2022optimization}.

The main goal of cloud portfolio management is usually achieving the lowest costs for running a specific set of applications over (a certain amount of) time.
Some research like Jangjaimon and Tzeng~\cite{jangjaimon2013effective} and Sharma et al.~\cite{sharma2017portfolio} tried to deal with this problem by creating a checkpointing mechanism or by focusing on preemptible servers in combination with concepts taken from financial modeling, to meet the requirements of applications when using spot instances. Meanwhile, Pittl et al.~\cite{pittl2019cost} took a more comprehensive approach to cloud portfolio management, which resulted in the findings that a more heterogeneous portfolio tends to be more cost-efficient. Another finding of that paper was to highlight the significance of right-sizing, where server instances should be chosen to most closely fit the capacity requirements of the applications the workload consists of. Otherwise, an expensive over-provisioning of resources could be the result~\cite{kiessler2022optimization}.
Regarding right-sizing, Hwang and Pedram~\cite{hwang2012portfolio} developed a portfolio-based optimization approach with a probabilistic model, which, while created for data center operations, is also applicable to cloud portfolio management. Each assigned workload in this model has a resource demand, not defined by a fixed value, but via a probabilistic model using a normal distribution. This idea of using probability-based problem formulation was also used in further research; for example, Martinovic et al. combined it with a stochastic bin-packing approach to finding efficient server allocations~\cite{martinovic2020stochastic}. One additional approach proposed for optimizing virtual machine placement to reduce energy consumption in data centers is using a genetic algorithm, as was done by  Wu et al.~\cite{wu2012energy}. Another interesting concept to be regarded is the one by De Cauwer et al., which introduced the idea of a temporal component to server allocation schemes while using deterministic resource demands. This additional dimension better reflects real computational needs, where applications often do not run forever but with a specific start and end time~\cite{de2016temporal,kiessler2022optimization}.

\subsection{Problem formulation}
\label{cpm_problem_formulation}
This section will deal with formulating the problem, which has to be tackled by the cloud portfolio optimization approaches. The notations used for this task are defined in \ref{tab:problem_notation}~\cite{kiessler2022optimization}.

\begin{table}
\begin{center}
\caption{\label{tab:problem_notation}Notation for problem formulation}

\begin{tabular}{cl}
 \toprule
 \multicolumn{1}{p{1cm}}{\centering \textbf{Parameter}} & 
 \multicolumn{1}{p{6cm}}{\centering \textbf{Description}} \\
 \midrule
 $I$ & A set of selected cloud instances (hosts) \\ 

 $A$ & A set of applications \\

 $T$ & A set of time slots for which to optimize \\
 
 $x_{ait}$ & Variable denoting assignments of applications to instances \\
 
 $S_a$ & The starting time of application $a$  \\
 
 $F_a$ & The finishing time of application $a$ \\
 
 $U_a$ & Indicates if application $a$ is preemptible \\

 $\mu_a$ & The expected resource demand of application $a$ \\
 
 $\sigma_a$ & The std. deviation of the resource demand of application $a$ \\

 $R_i$ & The resource capacity of instance $i$  \\

 $C_i$ & The cost of instance $i$ for one time slot \\

 $B_i$ & The first available time slot of an instance \\

 $E_i$ & The last available time slot of an instance \\
 
 $O_i$ & Indicates if instance $i$ is suitable for non-preemptible apps \\

 $D_{it}$ & Aggregated resource demand of instance $i$ at time $t$ \\
 
 $Q_{min}$ & The desired quality of service \\
 
 \bottomrule
 
\end{tabular}

\end{center}
\end{table}

First, we define a cloud portfolio as a set of \textit{cloud instances} $I$, which are used to run a set of \textit{applications} $A$ on them. The main goal of our optimization problem is to find a cost-efficient allocation of these applications and instances~\cite{kiessler2022optimization}.

All applications $A$ have a specific resource demand, which, as proposed in the literature by Hwang and Pedram, have their resource needs not defined as a fixed value but with fluctuation taken into account~\cite{hwang2012portfolio}. Therefore, we denote the capacity requirements of our applications as an expected demand mean $\mu_{a}$ and with a corresponding standard deviation $\sigma_{a}$. As our workloads also have varying run times, each application sports a starting time $S_a$ and a finishing time $F_a$, for which the statement $S_a < F_a$ must always be true. To model which applications are suitable for spot instances, meaning they can be interrupted at any time, we use the variable $U_a$ which can either be 0 or 1~\cite{kiessler2022optimization}. 

\begin{align}
U_a = \begin{cases}
1 & \text{if $a$ is preemptible} \\
0 & \text{else}
\end{cases}
\label{eqn:preemptible_applications}
\end{align}

For modeling instances, each instance $ I \in I$ has a predetermined resource capacity $R_i$, representing, for example, the number of CPUs and RAM of an instance. Furthermore, each kind of instance has a price per time unit $C_i$, where the total cost of any instance is calculated by the price per time slot and the overall up-time. As with applications, each instance has a given starting time $B_i$ and ending time $E_i$, with the inequation $B_i < E_i$ again having to be fulfilled at any time. Preemptible spot instances are denoted by the binary parameter $O_i$~\cite{kiessler2022optimization}.

\begin{align}
O_i = \begin{cases}
1 & \text{if $i$ is only suitable for preemptible applications} \\
0 & \text{else}
\end{cases}
\label{eqn:preemptible_instances}
\end{align}

To model the summed-up demand of all applications assigned to an instance $i \in I$, for a specific time slot $t \in T$, we use the parameter $D_{it}$. As with the demand of a single application, the aggregated demand is also not a fixed variable but a random variable, which means it can only evaluate the probability with which an instance stays within the designated capacity limits. Our proposed model also defines a desired minimum quality of service $Q_{min}$. An allocation is invalid if the probability of the aggregated demand $D_{it}$ staying below the provided capacity of the instance $R_i$ does not satisfy the minimum quality of service $Q_{min}$ for time slot $t$. This approach also builds upon the model proposed by Hwang and Pedram (2012). Still, it has to consider the temporal component of our model, meaning that the capacity restrictions have to be fulfilled for every time slot an instance is used~\cite{kiessler2022optimization}.

For modeling the formal assignment of applications to instances while considering the temporal restrictions of the problem, we used the approach postulated by Dell'Amico et al.~\cite{dell2020branch}. The variable $x$ denotes which hosting instance an application has been assigned to at a specific time. The resource demands for an application $a \in A$ are considered to be $0$ for any time slot $t \in T$ if $t < S_a$ or $t > F_a$~\cite{kiessler2022optimization}.

\begin{align}
x_{ait} = \begin{cases}
1 & \text{if app $a$ is assigned to instance $i$ at time slot $t$} \\
0 & \text{else}
\end{cases}
\label{eqn:assignment_variable}
\end{align}

Having outlined the cloud portfolio optimization requirements and assumptions, we can now present our exact problem statement. The main optimization goal is to find the minimum of the following cost function~\cite{kiessler2022optimization}:

\begin{equation}
    \label{eqn:min_portfolio_cost}
    min \sum_{i \in I} C_i * (E_i - B_i)
\end{equation}

\ref{eqn:min_portfolio_cost} is the main function to optimize. It minimizes the costs of the entire portfolio, which consists of the sum of all prizes incurred by each assigned instance. Hereby, $(E_i - B_i)$ is the time an instance is running, multiplied by its costs per time slot $C_i$ to arrive at the price the instance will account for. While trying to minimize \ref{eqn:min_portfolio_cost}, the following statements have to be fulfilled~\cite{kiessler2022optimization}:

\begin{align}
          \label{eqn:assign_app_to_instance}
    s. t. &\sum_{i \in I} x_{ait} = 1           &   &   &   &\forall a \in A, \forall t \in [S_{a}, F_{a}]\\
          \label{eqn:suitable_host}
          &\sum_{i \in I} x_{ait} * U_a \geq O_i  &   &   &   &\forall a \in A, \forall t \in [S_{a}, F_{a}] \\
          \label{eqn:resource_constraint}
          &P(D_{it} < R_i) \geq Q_{min}            &   &   &   &\forall i \in I, \forall t \in [B_{i}, E_{i}]\\
          \label{eqn:assignment_range}
          &x_{ait} \in \{0, 1\}                 &   &   &   & \\
          \label{eqn:preemptible_application_range}
          &U_a \in \{0, 1\}                     &   &   &   & \\
          \label{eqn:preemptible_instance_range}
          &O_i \in \{0, 1\}                     &   &   &   & \\
          \label{eqn:quality_range}
          &Q_{min} \in [0, 1]                     &   &   &   &
\end{align}

The first constraint \ref{eqn:assign_app_to_instance} asserts that each application has to be assigned to an instance, while at the same time, each, at any point of the run time of an application, can only be hosted by one instance. The next constraint \ref{eqn:suitable_host} states that each host has to come from a suitable market space, which is necessary to guarantee that non-preemptible applications are not assigned to spot instances, which could be interrupted at any time. The equation formulated in \ref{eqn:resource_constraint} ensures that for every time slot an instance is running, the probability of the resource demand of the applications assigned being within the capacity of said instance is at least the quality of service $Q_{min}$. The final four constraints \ref{eqn:assignment_range} to \ref{eqn:quality_range} state that the auxiliary variables have to be within a valid range from 0 to 1~\cite{kiessler2022optimization}.

\subsection{Optimization approaches}
\label{cpm_optimization_approache}
Having modeled our cloud portfolio optimization problem in \autoref{cpm_problem_formulation}, this section will now present our approaches to solving it. As our problem is essentially a multi-dimensional packing problem, it is NP-hard, very complex to solve, and finding an optimal solution is usually not computationally feasible~\cite{chekuri2004multidimensional}. Therefore, we developed two optimization heuristics to find good approximations of an optimal solution~\cite{kiessler2022optimization}.

\subsubsection{Greedy algorithm}
\label{cpm_greedy}
Our first algorithm is called \emph{Efficient Resource Inference for Cloud Hosting} (ERICH). It integrates the approach of the widely known bin packing algorithm first fit decreasing (FFD)~\cite{man1996approximation}, combining it with the proposed portfolio management strategy by Hwang and Pedram~\cite{hwang2012portfolio}. It is executed in the following four stages~\cite{kiessler2022optimization}:

\begin{itemize}
\item \emph{Stage 1}: As the first step, the algorithm sorts the preemptible and non-preemptible applications it receives as input by increasing starting dates, with applications that start earlier being allocated first. Applications with the same starting date are then sorted by non-increasing standard deviation of their resource demand, based on a proposal by Hwang and Pedram, suggesting that reduced capacity needs can be achieved by grouping workloads with similar resource demand deviation\cite{hwang2012portfolio}. To ensure that the algorithm prefers cost-efficient hosts, this step also includes sorting all received instance types by cost per time slot for the provided capacity in an increasing order~\cite{kiessler2022optimization}.

\item \emph{Stage 2}: Next, the algorithm tries to allocate all non-preemptible applications to reserved instances by using the first fit decreasing approach. Iterating through all such applications, the algorithm first tries to find a suitable host that provides the needed capacity over the entire run-time in the existing portfolio. If so, the application is assigned to the said instance. Otherwise, a new instance covering the needs of the application is added to the portfolio~\cite{kiessler2022optimization}.

\item \emph{Stage 3}: While reserved instances offer significant discounts in comparison to those procured on the on-demand market, the portfolio created in the previous step may be inefficient due to the requirements for minimum run-time in reserved instances. Therefore, step 3 tries to condense the portfolio by removing instances chosen in step 2 and replacing them with on-demand instances. To achieve this, the algorithm iterates through all instances, creating a new temporary portfolio with one reserved instance removed at each step. Next, applications from this instance are assigned to on-demand instances with the same first-fit-decreasing approach used in the previous step. Should this new allocation allow for a cheaper portfolio, it replaces the old one in the next iteration~\cite{kiessler2022optimization}.

\item \emph{Stage 4}: The final step of the algorithm deals with finding fitting instances for all preemptible applications, which can be assigned to multiple hosts during their lifecycle and can, therefore, be allocated on an individual timeslot basis. To leverage this, the algorithm will first find any time steps for a suitable candidate host and assign preemptible apps to these instances for the respective periods. Should there be any more need for the workload to run the apps, the difference is made up by adding new spot instances~\cite{kiessler2022optimization}.

\end{itemize}

Algorithm~\ref{alg:efficient_resource_inference_for_cloud_hosting} describes the steps discussed in pseudocode. To check if an application fits into any given instance, the equation \ref{eqn:resource_constraint} is used.

\begin{algorithm}
\caption{Efficient Resource Inference for Cloud Hosting}
\label{alg:efficient_resource_inference_for_cloud_hosting}
\SetAlgoLined
\SetKwInOut{Input}{Input}
\Input{A set of non-preemptible apps $A1$; A set of preemptible apps $A2$; A set of reserved instance types $RES$; A set of on-demand instance types $ON$; A set of spot instance types $SPOT$}
\KwResult{Packing pattern $portfolio$}

sort applications $A1$ and $A2$ by increasing start time and non-increasing $\sigma{_a}$\\
sort $RES$, $ON$ and $SPOT$ by non-increasing $C_i$ per $R_i$ and time slot\\
$portfolio$ $\gets$ empty allocation variable\\

\ForAll{$a \in A1$}{
    assign $a$ to $portfolio$ (FFD) while only considering $RES$ instances\\
} 

\ForAll{$i \in $ reserved instances from $portfolio$}{
    $tmp\_portfolio$ $\gets$ copy of $portfolio$ without instance $i$\\
    \ForAll{$a \in i$} {
        reinsert $a$ into $tmp\_portfolio$ (FFD) including $ON$ instances\\
    }
    \If{total cost of $tmp\_portfolio <$ total cost of $portfolio$}{
        $portfolio \gets tmp\_portfolio$\\
    }
}

\ForAll{$a \in A2$}{
    assign $a$ to $portfolio$ without allocating new instances\\
    $gaps \gets$ consecutive time slots where $a$ is not yet assigned to $portfolio$\\
    \ForAll{$gap \in gaps$}{
        assign $a$ to $portfolio$ for time slots in $gap$ by allocating $SPOT$ hosts\\
    }
}
\end{algorithm}

\subsubsection{Genetic algorithm}
\label{cpm_genetic}
Using a genetic algorithm (GA) to solve a bin-packing problem is not an entirely new idea. Still, it has been proven to work well in dealing with combinatorial optimization problems~\cite{reeves1996hybrid, falkenauer1996hybrid, quiroz2015grouping, kang2012hybrid}. GAs are based on genetic operators that can be adapted to fit a particular problem, enabling them to perform an efficient and targeted search in the problem space. Our genetic algorithm has been named \emph{Genetic Optimization of Resource Groupings} (GEORG) and will be described in this subsection. The algorithm is described in pseudocode in \autoref{alg:genetic_optimization_of_resource_groupings}, followed by a description of the algorithm building blocks~\cite{kiessler2022optimization}.

\begin{algorithm}
\caption{GEnetic Optimization of Resource Groupings}
\label{alg:genetic_optimization_of_resource_groupings}
\SetAlgoLined
\SetKwInOut{Input}{Input}
\Input{A set of non-preemptible apps $A1$; A set of preemptible apps $A2$; A set of reserved instance types $RES$; A set of on-demand instance types $ON$; A set of spot instance types $SPOT$}
\KwResult{List of packing patterns (portfolios) $population$}

$population$ $\gets$ use semi-random heuristic to create initial population\\
\While{termination criteria are not met}{
    $parents$ $\gets$ fitness-based selection of individuals from $population$\\
    $offspring$ $\gets$ apply temporal biased crossover for each tuple in $parents$\\
    $offspring$ $\gets$ repair broken chromosomes in $offspring$ after crossover\\
    $offspring$ $\gets$ apply domination mutation operator to random $offspring$\\
    $offspring$ $\gets$ repair broken chromosomes in $offspring$ after mutation\\
    $population$ $\gets$ fitness-based merge of $offspring$ and current $population$\\
}
\end{algorithm}

\begin{itemize}
\item \emph{Encoding scheme.} Grouping of items and using bins are essential for building a GA, according to Falkenauer~\cite{falkenauer1996hybrid}. Their approach of encoding, where each chromosome consists of an array of bins holding a set of items, is not suitable to our problem with its temporal component. That is why we have chosen a temporal group encoding, where every chromosome locus represents a certain time step, while the allele is a set of instances running at a certain time. Finally, every host is assigned a set of applications, which are then allocated to the corresponding instance during this time slot.~\cite{kiessler2022optimization}.

\item \emph{Population initialization.} To enable the operations of a GA, like crossover, mutation, and survivor selection, to occur, an initial set of individuals (a population) is needed. For our initialization process, a hybrid approach was chosen to achieve comparatively high fitness from the start while also offering good genetic diversity. Therefore, half of the assignments are done randomly, with the other half having applications that have been assigned to reduce the number of allocated instances and the overall costs~\cite{kiessler2022optimization}. 

\item \emph{Fitness evaluation.} This building block of our GA uses the equation \ref{eqn:min_portfolio_cost}, the main function to optimize, to evaluate the fitness of each individual~\cite{kiessler2022optimization}.

\item \emph{Parent selection and crossover.} For each new generation of the GA, a group of individuals is chosen for parentage of the following generation. These individuals are chosen based on the fitness proportionate roulette wheel method. Out of these, there is a crossover applied by pairs of two parent solutions to create new solutions (offspring). To perform this crossover, a new biased temporal crossover operator was built on the concepts proposed by Quiroz-Castellanos et al.~\cite{quiroz2015grouping}, which aims to encourage the passing on well-fitting genetic material to the following generation. Within each gene, which represents a time step, all active instances are sorted by decreasing the average capacity utilization rate and cost per time slot. Using a zip-merging approach, a new partial solution is created from both parent solutions. After removing hosts of already assigned applications, the partial solution is pruned. If any instances violate the constraints outlined in the problem formulation, they will be pruned in the subsequent time steps of the crossover process. Culling instances from solutions may break some chromosomes, as applications can end up with no or only partial assignments. This issue is addressed by employing a basic heuristic to reintroduce any applications that have not been completely assigned~\cite{kiessler2022optimization}. 

\item \emph{Mutation.} This operator is used on freshly produced offspring randomly to introduce new genetic characteristics to individuals and enhance the population's overall fitness. The concept of dominance, introduced by Martello and Toth~\cite{martello1990lower}, can lead to tighter packing patterns by replacing a subset of items with an item of larger or equal size. This approach has been shown to be incorporable into the mutation operator of a GA~\cite{falkenauer1996hybrid, quiroz2015grouping}, though for usage in this work, the definition of dominance has to be adapted. Application $a$, hosted by instance $i \in I$ for the time slots $[S_t, F_t]$, dominates a partition of apps $P$ from instance $i$ if the period denoted by $[S_t, F_t]$ contains all assignments slots of the partition for the respective host. Additionally, the probability of the resource demand of the dominating application being higher than the summed-up capacity requirements of all elements of partition $P$ has to exceed fifty percent. Applying the mutation operator on an individual results in the removal of several instances from the portfolio. Each application that is now unassigned is checked against candidate hosts from the portfolio. By creating partitions of applications of size two from the respective candidate instance, we can try to find partitions dominated by the unassigned applications. Should one be found, the application is swapped with the partition. As these operations may leave broken chromosomes with unassigned applications, just like the crossover operator, the insertion heuristic mentioned previously is used again in order to repair those corrupted individuals~\cite{kiessler2022optimization}. 

\item \emph{Insertion heuristic.} One of the main constraints of our problem formulation is equation \ref{eqn:assign_app_to_instance}, which requires each application to be assigned to a fitting instance at any time slot of its run time. As previously mentioned, the crossover and mutation operators may violate said constraint. The insertion heuristic chosen to alleviate this problem uses a naive first-fit approach, which chooses a candidate instance from the existing chromosome to fit orphaned applications. Should a non-fitting host exist in the portfolio, a new random instance is generated to accommodate the corresponding application. The element of randomness is used to reduce the risk of converging on a local optimum by creating additional genetic diversity~\cite{kiessler2022optimization}.

\item \emph{Survivor selection.} The process of survivor selection decides which individuals at the end of an iteration will represent the next generation. This can be achieved in several ways, with one of the simplest being the selection of the fittest individuals. While more sophisticated methods exist, and simply choosing the best individuals could lead to a lack of diversity and converging on local optima, both our crossover and mutation operators introduce enough randomness, meaning that the fittest individuals are sufficient for our algorithm. 

\item \emph{Termination.} For termination, our GA can use one of several common stopping criteria, like a maximum number of generations, the fitness scores of the individuals of a population converging to a certain degree and therefore becoming very similar, or the highest level of fitness of an individual not changing any more with more generations~\cite{safe2004stopping, kiessler2022optimization}.

\end{itemize}

\subsection{Evaluation and Results}
\label{cpm_evaluation_results}
In this section, we will discuss how the previously described algorithms were evaluated and the results of this evaluation to present what kind of optimization results a customer could expect from the cloud portfolio optimizer. Further examples created in the Cloud Portfolio Manager platform will be presented in \autoref{cloud_portfolio_manager}.
To evaluate our optimization heuristics, we used synthetic data. The implementation of the algorithms was done in Python 3.9, and the tests were conducted on a PC running a Windows operating system with an Intel Core i7-4770 processor (3.4 GHz base clock, 3.9 GHz turbo) and 16 GB of DDR3 memory at 1600 MHz~\cite{kiessler2022optimization}.

\subsubsection{Data set description}
\label{data_set_description}
With our optimization problem, multiple dimensions influence the difficulty of any test set. As is the case with any bin packing problem, a primary contributor to this is the number of items to be assigned. Unlike many other bin packing problems, our problem must regard the temporal component as the primary driver of execution time, with the number of allocation periods increasing the difficulty. Therefore, the created test data sets reflect a variety in both these attributes~\cite{kiessler2022optimization}.

\begin{table}[H]
\caption{\label{tab:application_data}Summary of application data sets}
\begin{center}
\begin{tabular}{lrrrrrrrrr}

\toprule
\multicolumn{1}{p{0.3cm}}{\centering \textbf{App. \\ Set}} &  
\multicolumn{1}{p{0.3cm}}{\centering \textbf{Non-Pre.}} &
\multicolumn{1}{p{0.3cm}}{\centering \textbf{Pre.}} &
\multicolumn{1}{p{0.3cm}}{\centering \textbf{Avg. \\ Res. \\ Dem.}} &
\multicolumn{1}{p{0.3cm}}{\centering \textbf{Std. \\ Res. \\ Dem.}} &
\multicolumn{1}{p{0.3cm}}{\centering \textbf{Avg. \\ Res. \\ Dev.}} &
\multicolumn{1}{p{0.3cm}}{\centering \textbf{Std. \\ Res. \\ Dev.}} &
\multicolumn{1}{p{0.3cm}}{\centering \textbf{Avg. \\ Alloc. \\ Periods}} &
\multicolumn{1}{p{0.3cm}}{\centering \textbf{Std. \\ Alloc. \\ Periods}} &
\\
\midrule
apps\_1 & 14.0 & 6.0 & 3.2 & 1.7 & 0.5 & 0.5 & 43.1 & 33.4 \\
apps\_2 & 59.0 & 41.0 & 3.0 & 2.6 & 0.5 & 0.7 & 63.9 & 43.9 \\
apps\_3 & 10.0 & 10.0 & 3.0 & 2.0 & 0.7 & 0.6 & 212.2 & 167.8 \\
apps\_4 & 42.0 & 58.0 & 3.1 & 2.6 & 0.5 & 0.6 & 237.2 & 171.5 \\ 
apps\_5 & 7.0 & 13.0 & 3.1 & 2.7 & 0.6 & 0.6 & 2758.5 & 1996.9 \\
apps\_6 & 41.0 & 59.0 & 2.8 & 2.0 & 0.5 & 0.6 & 2871.7 & 2055.6 \\

\bottomrule

\end{tabular}

\end{center}

\end{table}

Even though the test data used for the evaluation was synthetically created, it considers realistic scenarios, including anticipated price discounts for spot and reserved instances compared to on-demand instances. This was based on observations done on the major cloud service providers AWS, Google Cloud, and Microsoft Azure. 
Additionally, for creating our test instances, the price-to-capacity ratio has been modeled similarly to offerings observed on the previously mentioned CSPs. The table \ref{tab:application_data} above shows our chosen six sets of applications with their key resource demands and allocation characteristics, while the table \ref{tab:instance_type_data} describes the three sets of instance types created for testing, each containing 500 instance types. These were combined in the following pairings to create six test cases: $case\_1$ ($apps\_1$, $types\_1$), $case\_2$ ($apps\_2$, $types\_1$), $case\_3$ ($apps\_3$, $types\_2$), $case\_4$ ($apps\_4$, $types\_2$), $case\_5$ ($apps\_5$, $types\_3$) and $case\_6$ ($apps\_6$, $types\_3$)~\cite{kiessler2022optimization}.

\begin{table} [H]
\begin{center}
\caption{\label{tab:instance_type_data}Summary of instance type data sets}
\begin{tabular}{lrrrrrrrrr}
\toprule
\multicolumn{1}{p{0.4cm}}{\centering \textbf{Inst-ance \\ type \\ set}} &  
\multicolumn{1}{p{0.4cm}}{\centering \textbf{Avg. \\ capa-city}} &
\multicolumn{1}{p{0.4cm}}{\centering \textbf{Std. \\ capa-city}} &
\multicolumn{1}{p{0.4cm}}{\centering \textbf{Avg. \\ Res. \\ Prc.}} &
\multicolumn{1}{p{0.4cm}}{\centering \textbf{Std. \\ Res. \\ Prc.}} &
\multicolumn{1}{p{0.4cm}}{\centering \textbf{Avg. \\ On. \\ Prc.}} & 
\multicolumn{1}{p{0.4cm}}{\centering \textbf{Std. \\ On. \\ Prc.}} &
\multicolumn{1}{p{0.4cm}}{\centering \textbf{Avg. \\ Spot \\ Prc.}} &
\multicolumn{1}{p{0.4cm}}{\centering \textbf{Std. \\ Spot \\ Prc.}} &
\\
\midrule
types\_1 & 9.6 & 8.8 & 2.3 & 2.8 & 3.1 & 2.2 & 2.5 & 2.2 \\
types\_2 & 10.3 & 11.4 & 2.2 & 2.4 & 3.1 & 2.6 & 3.1 & 4.8 \\
types\_3 & 9.8 & 9.9 & 2.4 & 3.8 & 3.1 & 2.4 & 2.3 & 1.7 \\

\bottomrule
\end{tabular}
\end{center}
\end{table}

\subsubsection{Results}
\label{cpm_results}
The results of our tests focus on three key criteria for evaluation: speed of execution, packing density, and overall costs incurred by the resulting portfolio. To avoid side effects other tasks running on the test machine may have, each algorithm ran ten times to get data on execution speeds.

\begin{figure}[ht]
    \centering
    \includegraphics[width=0.95\linewidth]{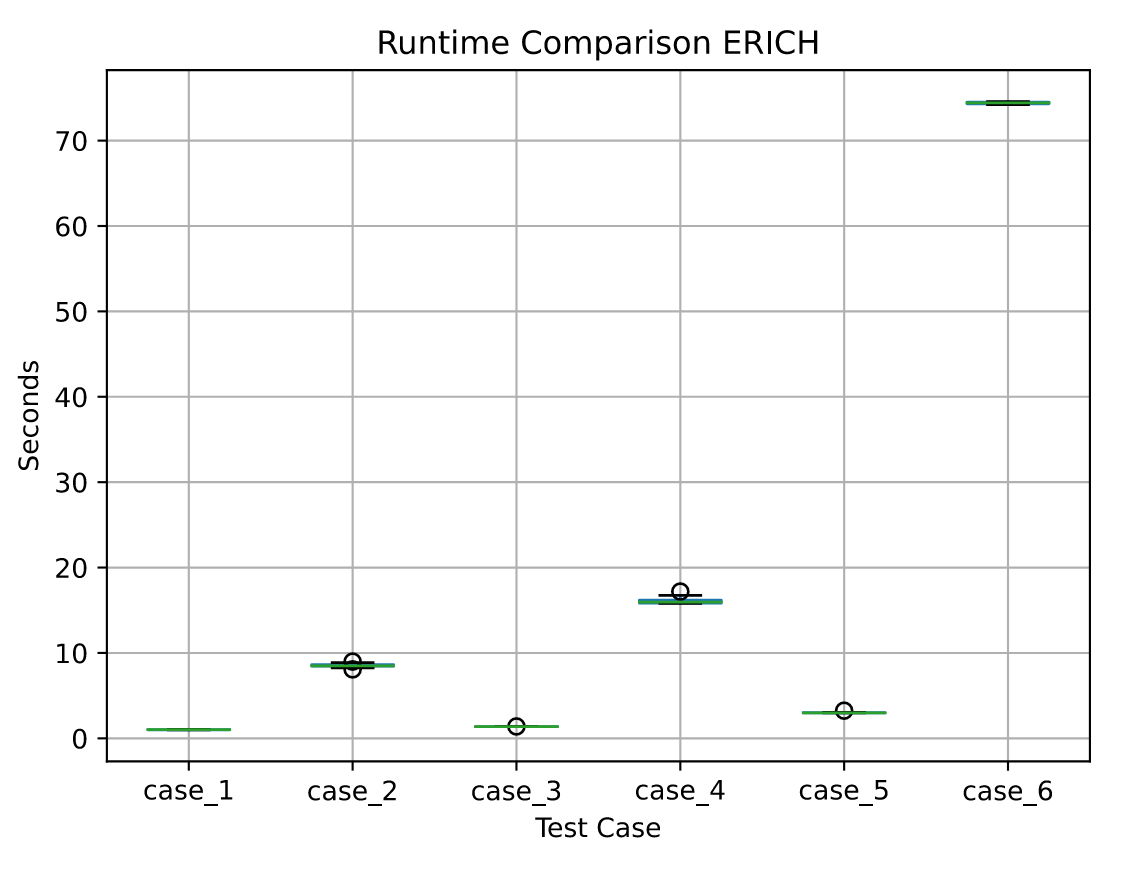}
    \caption{Execution time ERICH}
    \label{fig:cpm_results_speed_erich}
\end{figure}

\begin{figure}[ht]
    \centering
    \includegraphics[width=0.95\linewidth]{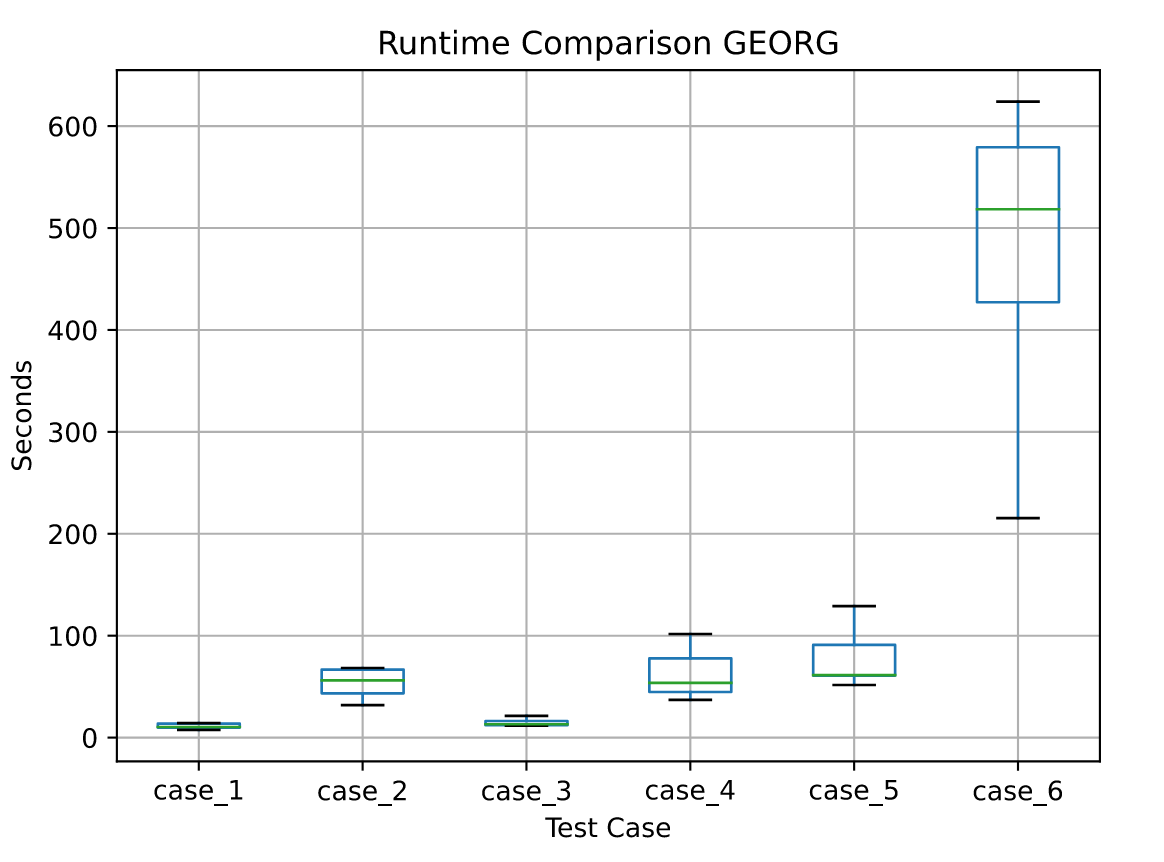}
    \caption{Execution time GEORG}
    \label{fig:cpm_results_speed_georg}
\end{figure}

As you can see in \autoref{fig:cpm_results_speed_erich} and \autoref{fig:cpm_results_speed_georg}, the optimization approach \emph{ERICH} offered way faster execution speeds, being faster by the magnitude of close to 10. It also results in an almost static execution speed by being deterministic. In contrast, the optimization approach \emph{GEORG} is not only slower, but also way more volatile in terms of execution speed. The slowest run can take 2-3 times longer than the fastest one due to genetic operators being highly influenced by randomness~\cite{kiessler2022optimization}.

\begin{figure}[ht]
    \centering
    \includegraphics[width=0.95\linewidth]{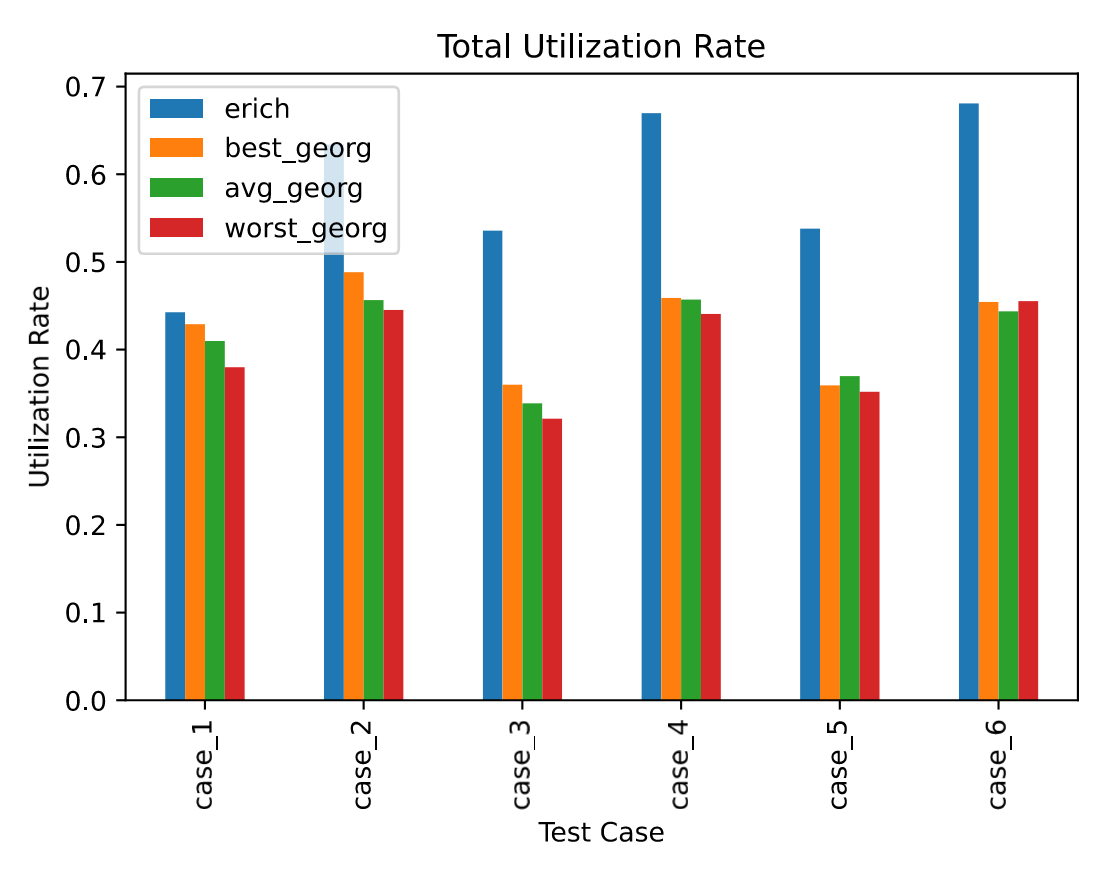}
    \caption{Utilization comparison between ERICH and GEORG }
    \label{fig:cpm_results_util_total}
\end{figure}

Our second criterion, the utilization rate, measures the relationship between the total expected resource demand of all assigned applications across relevant time slots and the absolute capacity provided for that period. Depicted in \autoref{fig:cpm_results_util_total}, it is easy to see that once again \emph{ERICH} delivers better results than \emph{GEORG}. Depending on the test case, the gap can range from minor, like in case 1, to very significant, as in cases 4 and 6~\cite{kiessler2022optimization}.

\begin{figure}[ht]
    \centering
    \includegraphics[width=0.95\linewidth]{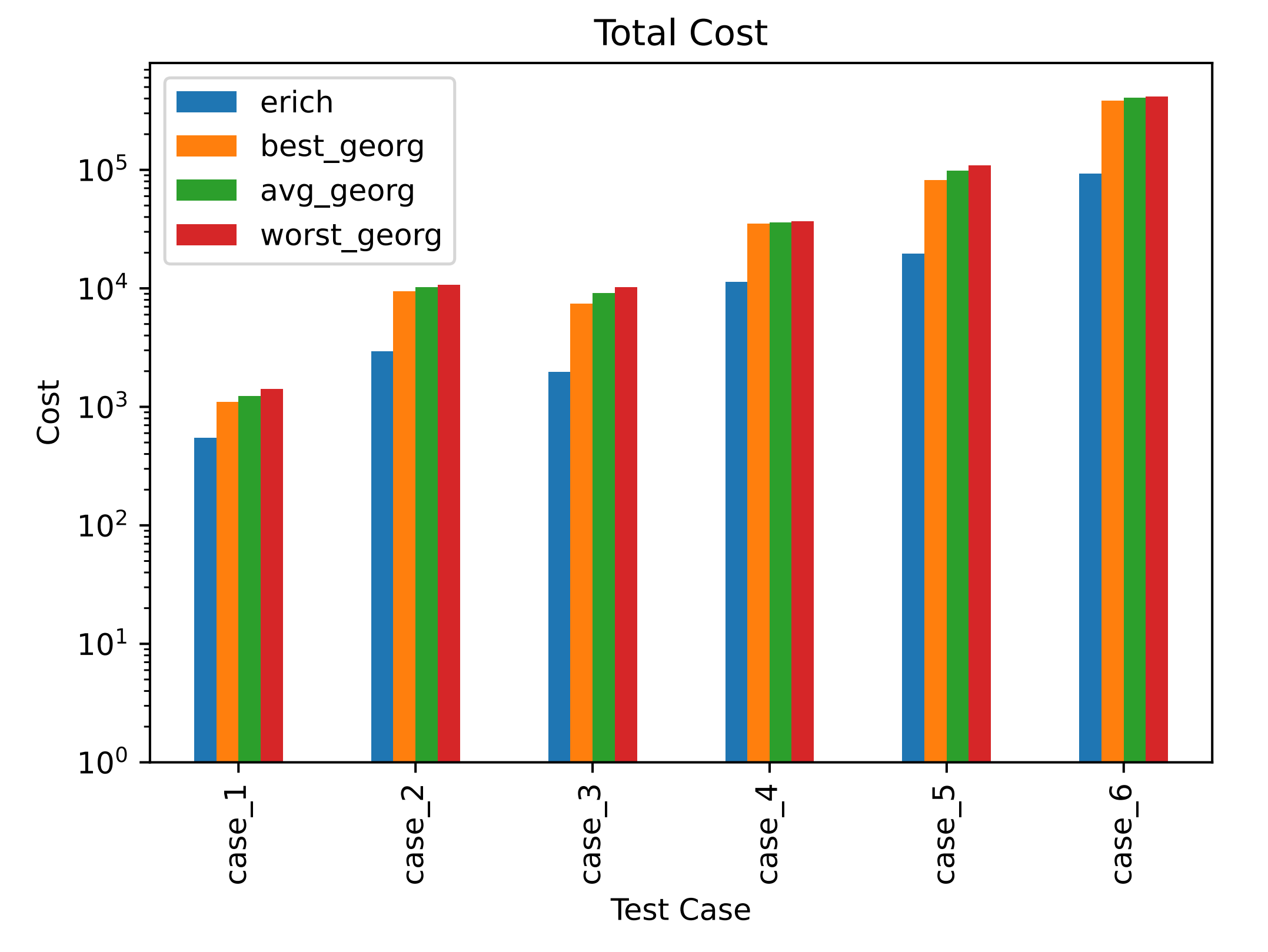}
    \caption{Portfolio comparison between the two algorithms}
    \label{fig:cpm_results_costs_overall}
\end{figure}

\begin{figure}[ht]
    \centering
    \includegraphics[width=0.95\linewidth]{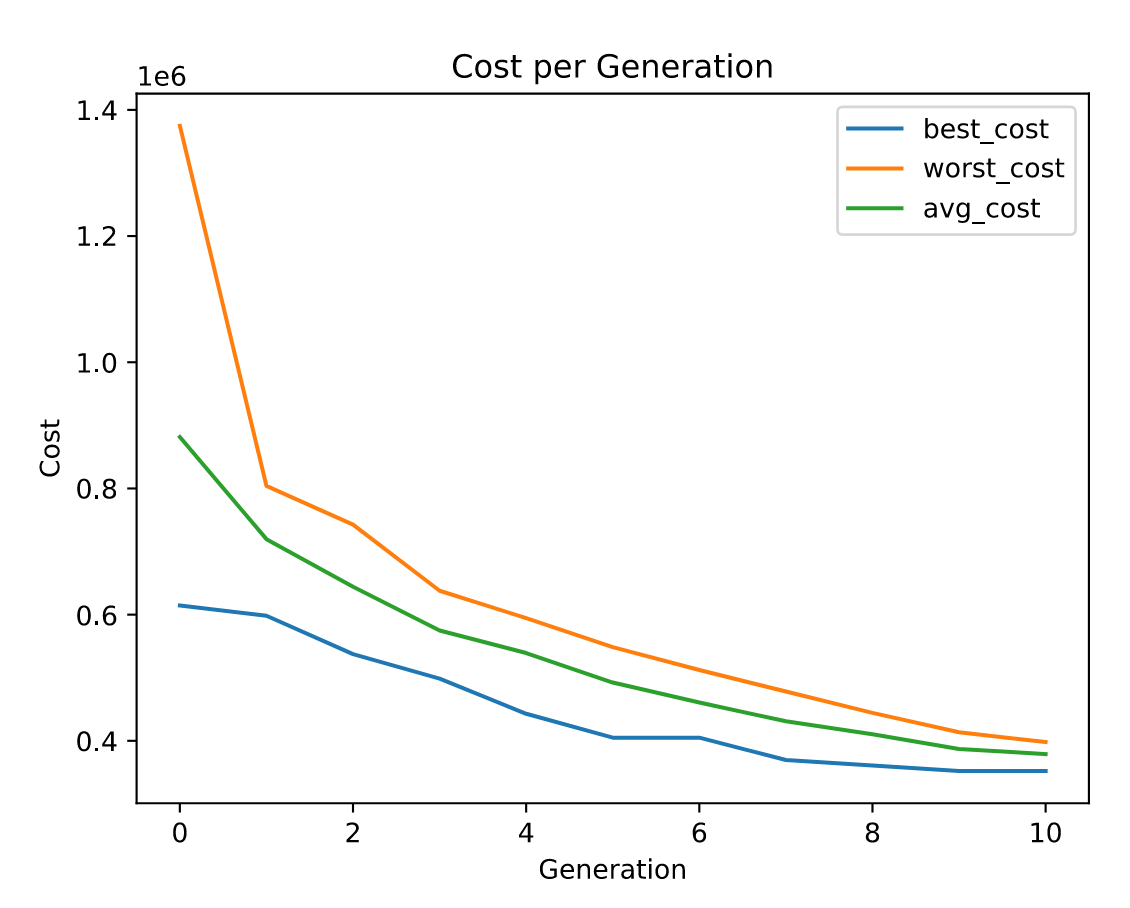}
    \caption{Cost for each generation for set $case\_6$ using GEORG}
    \label{fig:cpm_results_costs_trend}
\end{figure}

Finally, in terms of overall costs of the generated portfolio, \emph{ERICH} also outperformed the GA by a significant margin, which can be seen in \autoref{fig:cpm_results_costs_overall}. While this may initially lead to the conclusion that the GA did not work correctly, this is untrue. In \autoref{fig:cpm_results_costs_trend}, one can observe that the costs, meaning the fitness level of the multiple generations for the data set $case\_6$ improve continuously, with each new generation being fitter than the previous one. The initial population starts with a high degree of genetic diversity and improves with each generation. After ten generations, which is the duration the test ran, the average costs have been reduced by more than 50\%. As the initial population of the GA can serve as an example of how an unplanned allocation of resources would look like, it also shows that both algorithms can deliver notable lower costs for a portfolio and, therefore, offer a usable basis to build our Cloud Portfolio Manager platform upon~\cite{kiessler2022optimization}.  

\section{Business Model}\label{buiness_model}

In this section we will first present our business model of the Cloud Portfolio Manager using the Business Model Canvas framework by Osterwalder and Pigneur~\cite{osterwalderbusinessmodel}, where we describe our model with the help of the nine building blocks introduced in~\autoref{state_of_the_art}). Afterwards, we will discuss how our model can be classified within the context of various classification approaches. Finally, we will compare our model to contender platforms offering services with some similarity to our own.

\subsection{The Cloud Portfolio Manager Business Model Canvas}

\subsubsection{Customer Segments}
For our customer segment, the sole focus will be on the B2B area, as consumers, so far, have little to no reason to purchase cloud resources for personal use. Within the B2B sector, a big emphasis will be put on the IT sector, ranging from small businesses to large companies, seeing as large parts of this sector move more and more of their IT infrastructure into the cloud. One example would be Netflix, which, since 2015 has moved its entire IT infrastructure to AWS~\cite{amazon_netflix}.

Surprisingly, when it comes to smaller businesses, as suggested in a study by Jonas et al. in 2013, start-up companies looking for cloud solutions prefer the reputation of a cloud provider over other aspects such as price, security, and reliability~\cite{cloud_startup_preferences}. At least at the beginning, a small and not well-known company could mean targeting this customer segment could be more challenging than others. This situation also leads us to conclude that establishing a reputation should be a primary business goal. Besides this, we can increase our attractiveness to small businesses and start-ups by offering cloud consulting and optimization services. These services aim to help customers better understand cloud environments and set up and operate their own cloud portfolios. While it can be expected that larger companies already have access to this knowledge, the same cannot be said for smaller businesses. Aided by our consultancy services, we can also market our optimization service to this customer segment.

One customer segment our business will specifically focus on within the IT sector is those companies, businesses, and possibly research facilities working with machine learning algorithms. These are uniquely well-suited to be run on cloud resources, as their implementations often offer out-of-the-box preemptibility like PyTorch and Google TensorFlow~\cite{pytorch_checkpoints,tensorflow_checkpoints}. Combining machine learning algorithms, resource-intensive applications, and our optimization approach, delivering the best results for preemptible applications, makes them a perfect match. Our value proposition, which will be detailed within its own building block, is well suited to reduce one of the great pains of implementing machine learning solutions, the lack of processing power, by offering cheaper access to computational resources.

While the previously described customer segment will be our primary focus, other sectors also offer a wide range of potential customers. Even in 2017, Mohit et al. suggested that over 90\% of organizations were either already adopting cloud infrastructure or planning to do so within the next one to three years~\cite{cloud_market_predictions}. Therefore, our cloud resource optimization approach's potential market is large and diverse.

\subsubsection{Value Proposition}
Our primary value proposition, a cost reduction for their cloud portfolio, is unlike many other business models, targeting all customer segments. It can be applied to both existing portfolios and first-time cloud deployments, as long as the customer is roughly aware of their application's computational needs and run-time. A portfolio can be set up in two ways: First, directly via interacting with the platform via its website. Alternatively, outside initial registration, most interactions with the platform can also be done via API, allowing customers to integrate our service within their own systems and automate the process of creating portfolios and creating allocations.

The previously mentioned cost reduction is achieved by a combination of choosing the cheapest instances from the right marketplace and reducing their idle time through continuous monitoring of the needs of the applications. The resulting benefit for the customer entails the direct cost reduction itself and offers easier access to the complex world of cloud computing, which could be especially useful for small and medium-sized businesses. For these clients in particular, we can extend our value proposition by offering a consultancy service for customers who still need the know-how required to take advantage of cloud solutions. Said service would focus on the basics of cloudification, such as which applications are viable for being put into the cloud, the creation of portfolios, and first-time deployment.

\subsubsection{Channels}
Listed here, we will address the channels used throughout the five phases of customer interaction through which we aim to reach our customers:

\begin{itemize}
    \item \emph{Awareness:}
    For the first phase, raising the customer's awareness of our service, a mixture of online advertisement and direct contact with potential customers through an in-house sales force seems applicable. Furthermore, targeted online advertisement, for example, via Google ads\footnote{\url{https://ads.google.com/intl/de_at/home/}}, could be considered an option. While having the potential to result in a higher click-through rate, the advertiser has to be careful not to be too intrusive, as this can result in having the opposite effect. It also appears that targeted advertising does not work well with every demographic~\cite{targeted_ads_survey}. After the first customers have been acquired, it can also be reasonably expected that word-of-mouth between different businesses could raise further awareness levels for the company. Another option to reach potential customers is trade fairs focusing on the IT sector.  

    \item \emph{Evaluation:}
    Our website is the primary channel used for evaluation and the center of the operation. It provides example calculations, which showcase the potential cost reductions offered by the service, gives an overview of the pricing structure, and offers guidance on how to set up an account. Later, customer success stories, a widely adopted practice among online businesses today, will be showcased on the website. These stories further highlight tangible implementations of our service and provide credibility to our claims of assisting customers in optimizing their cloud portfolio.

    \item \emph{Purchase:}
    Regarding purchasing our products, many online payment services such as PayPal, Amazon Payments, and Credit cards are available and can be provided with relative ease. Besides directly integrating single payment options, companies such as Stripe\footnote{\url{https://stripe.com/en-gb-at/payments}} offer a single API that enables the user to choose from a range of standard online payment options.

    \item \emph{Delivery:}
    The delivery of the optimization service to the customer can also be achieved through the platform's direct channel, either by direct customer interaction on the website or via API call. As for the consultancy side of the business, we expect delivery to be done via personal interaction, both physical and online, depending on the customer's preferences.

    \item \emph{After sales:}
    The Cloud Portfolio Manager will first focus on providing post-purchase support to customers through our website. Here, the user can overview his portfolios, optimizations, subscriptions, and general account information. A FAQ page can also help answer general questions. Direct support through a personal customer support force should also be implemented for more complex cases.
\end{itemize}

\subsubsection{Customer Relationships}
When it comes to the methods of interacting with the customer base, the Cloud Portfolio Manager will focus on two areas: For the majority of interactions, such as setting up an account, creating and managing a portfolio, an automated service based on our website will be used. Complementing these services is a sales- and CRM- (Customer Relationship Management) force, which can be contacted personally via channels such as e-mail, phone, or video calls. This enables a more personal relationship with the customer and answers complicated and personal questions regarding single customers, which cannot be served easily via an automated service. This additional service is available to the customer during the whole interaction, from pre-sale evaluation until the purchase is completed. Another important aspect of this building block is the interaction between the consultancy force and the customers. As for offering a consultation service, direct human interaction is preferable, as each customer is assigned and mainly interacts with one consultant. This can be over various channels, but due to the close nature of the relationship, it will result in more face-to-face interactions than the other services of our business. A study by Roy et al. suggests that direct interaction with the customer is also preferable, as service experience is valued even higher than the actual service quality in B2B services~\cite{service_quality}.

\subsubsection{Revenue Streams}
Regarding revenue streams, a plethora of options are available at first glance. However, as we state in this section, most of them are not readily applicable to our platform for one reason or another, leaving us with one very widely used revenue stream as our primary source of revenue.

The first option we want to discuss is advertisement. While it is the main revenue stream of many large online platforms such as YouTube and Facebook, these are mass media B2C operations with millions or even billions of users, where each user only generates a relatively small amount of revenue through displayed advertisements. For our platform, which offers a specialized service to business customers, advertisements would mainly discourage users and possibly damage the brand reputation~\cite{revenue_cloud_platforms}.

Next, we have considered the option of transaction fees. This could be implemented on a usage-based model, where the customer would pay a certain amount for each optimization based on the portfolio size. Another implementation could be a one-time transaction enabling unlimited access to the Cloud Portfolio Manager. Both options are not optimal for our product. The usage-based model does not lend itself well to a product that is meant for continuous optimization. This would either lead to a need to frequently pay for a new allocation or prevent customers from getting the full benefit of an approach that is meant to adapt to changing demands in their portfolio. The one-time charge option faces another drawback, making it unfeasible. Seeing as our portfolio manager is intended as a continuous service, this one-time charge would have to account for a long service time, which in turn would increase the price a level, which would turn it into a severe deterrent for new customers that are not entirely convinced of the benefits of the product yet.

Another alternative would be a system based on a brokerage fee. In this case, a part of the cost reduction achieved by the Cloud Portfolio Manager would be taken as our revenue. The significant flaw with this idea, though, is that our system does not aim to directly access and manage the customer's cloud instances. This would result in customers needing to accurately and honestly report their current cloud expenses and their achieved cost reductions, which lends itself to be abused way too easily.

Finally, we propose that subscription fees are the revenue stream best suited to our business model. They tackle several disadvantages mentioned in the previously discussed systems, such as fitting well with a continuously running service, unlike pay-per-use transaction fees and a low entry barrier compared to a one-time charge. The subscription fee, due in a monthly interval, could either be based upon a system with different levels of subscriptions, offering support to differing sizes of cloud portfolios and varying levels of customer support, or directly scaling with the size of the optimized cloud portfolios.

As for revenue streams concerning the consultancy side of the business model, three monetization variants are possible. First, a classic hourly fee would most suit customers needing only a more minor assistance contingent. Another variant would be offering package deals with a fixed price, such as offering to help set up the first cloud portfolio for a customer. Finally, higher-level subscription models for the cloud portfolio optimizer platform could include a certain amount of consultancy services for free.

\subsubsection{Key Resources}
As for the differing categories of key resources, the following can be said: When it comes to physical key resources, there is little to be mentioned here. While server resources are necessary to host the platform, the hardware required is highly interchangeable and easy to come by. Furthermore, it could be more advantageous to completely forgo physical servers and host the platform itself on a cloud server.

Financial key resources may also not play a huge role in starting off. Of course, financial resources such as cash or credit will be needed to set up the business, but due to its nature, these will be of a small volume. One possible option to gain access to financial resources to start the business would be to apply for one of the many tech start-up sponsorships available in Austria. The most critical key resources are within the intellectual resource category, consisting of the portfolio management platform and, in particular, the optimization algorithms, which are at the heart of the operation and are needed to realize all of the other components of the business model.

Finally, when it comes to human key resources, the following groups can be expected to be part of those: Especially for development and improvements to the platform, further full-stack developers could be needed. Furthermore, a small team of cloud consultants would be responsible for providing customers with know-how on cloud solutions. Besides that, a group of employees helping with sales and CRM-related topics should be employed as well.

\subsubsection{Key Activities}
The most important key activity of the business is the operation and maintenance of the Cloud Portfolio Manager platform. In this capacity, the platform offers the customer an automated service. After creating an account and logging in, customers can create, delete, and change their cloud portfolios. Besides a simple interface to directly manipulate a portfolio, the main feature for management is the possibility to upload load profiles based upon which an optimized portfolio of instances is calculated and displayed to the customer. The load profiles can be uploaded manually on the website and through a REST API.

In addition to the service provided by the platform, the other main activity is problem-solving for the customer by offering our consultancy service. This mainly entails sharing cloud-related know-how and aiding the customer with planning, creating, and monitoring their own cloud solutions and migrating their existing applications.

\subsubsection{Key Partnerships}
Within this building block, the most prevalent partnerships are the various CSPs for which the platform offers portfolio optimization. While actively managing the customer's portfolio is not part of the business plan so far, it is crucial to the platform's functionality to access the instances and their respective pricing offered by the various providers. Luckily, all major CSPs provide APIs that give access to live data on the current availability and pricing of their offered instances.

If the business grows beyond a small scale, it would be possible for certain activities, such as customer support or cloud consulting, to be outsourced to external partners, which would turn these into key partnerships as well.

\subsubsection{Cost Structure}
The cost structure of the business model is intended to lean towards being value-driven, focusing on creating value for the customer. As the business operates online with a web platform at its center, scaling should be achievable relatively easily. While an increase in the customer base will require additional staff for CRM and consulting, the platform performance for handling a certain amount of customers can be scaled almost infinitely and with a mediocre but easily projectable impact on costs.

In contrast to the platform's operating costs, which should be manageable, the same cannot be said about its initial creation, which can be expected to be one of the significant cost factors in starting the business. Once the platform is in operation, the major cost factors will be the personnel required for the CRM, consulting operations, and resources spent on updating and expanding the platform. Further costs that have to be taken into account come from actions taken towards the acquisition of customers, especially those mentioned in the awareness section of the "channels" building block.

\begin{figure}[ht]
    \centering
    \includegraphics[width=\linewidth]{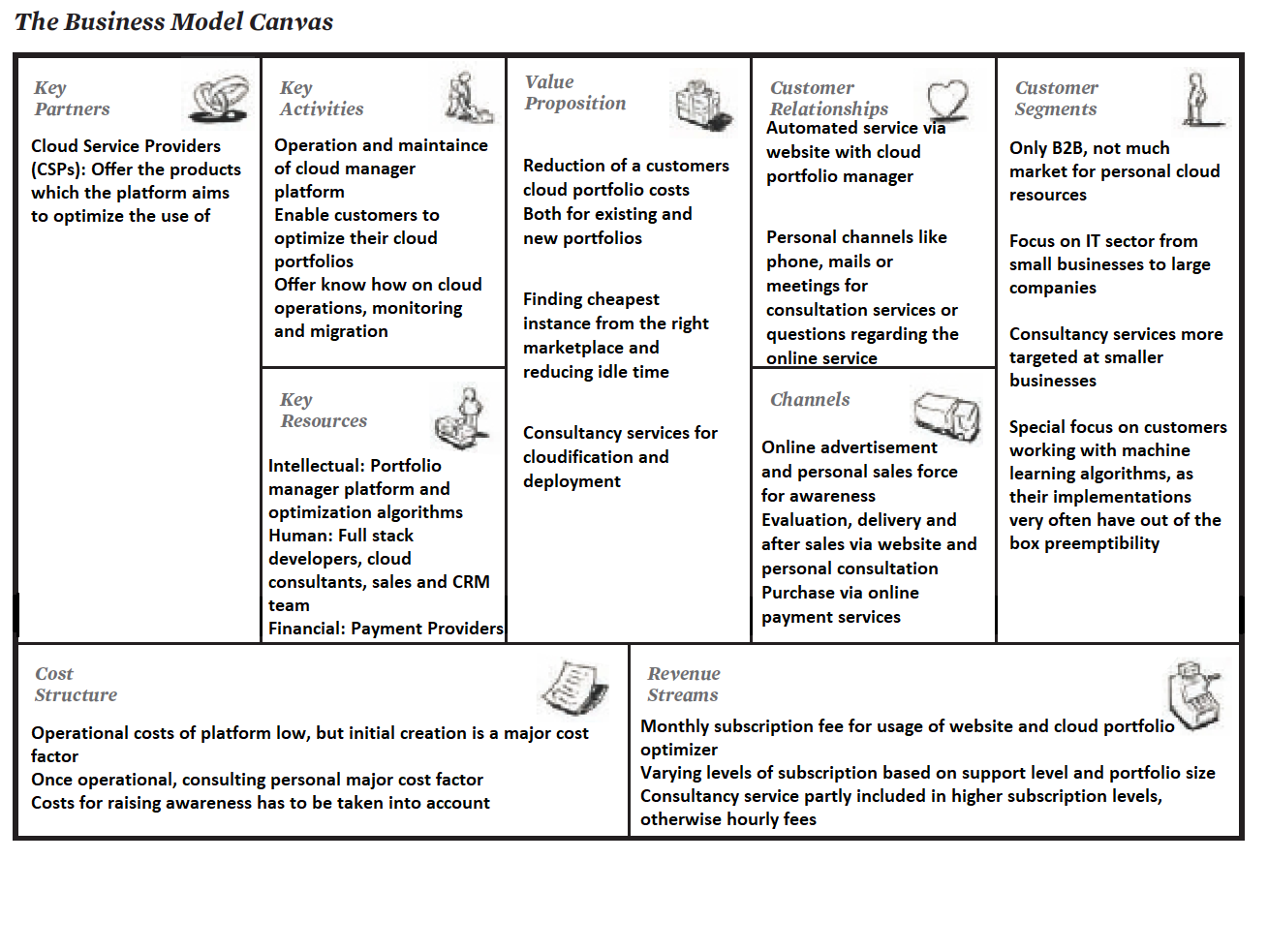}
    \caption{Business Model Canvas~\cite{osterwalderbusinessmodel} of Cloud Portfolio Manager}
    \label{business_model_canvas_filled_out}
\end{figure}

\subsection{Model Classification}

Having described the Cloud Portfolio Managers business model, we will now discuss how our model can be classified within various classification systems developed in the literature. Besides the system offered by Timmers~\cite{timmers1998business}, we will also apply the classifications of Osterwalder and Pigneur~\cite{osterwalderbusinessmodel}.

\subsubsection{Classification according to Timmers}
\begin{figure}[ht]
    \centering
    \includegraphics[width=\linewidth]{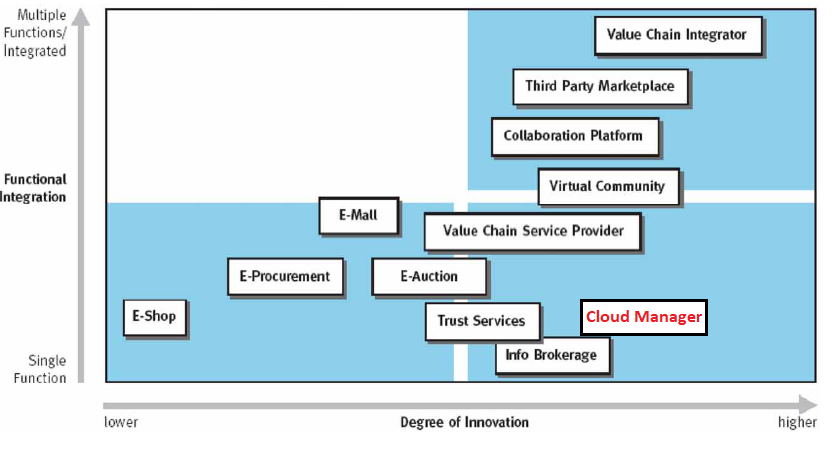}
    \caption{Classification of Cloud Portfolio Manager within internet business models, figure from B. Wall et al., 2007, Production Planning and Control, page 248~\cite{review_business_model}}
    \label{classification_buiness_model_placement}
\end{figure}

Many authors worked on classifications for e-business models like the classification made by Timmers~\cite{timmers1998business}, which, despite the age of the paper, still applies well to today's e-business models. The classes are the E-shop, E-procurement, E-auction, E-mall, third-party marketplace, virtual communities, value-chain service provider, value-chain integrator, collaboration platform and information broker, trust, and other services. These different models are all aligned along two criteria: functional integration, from a single function to multiple, and their degree of innovation from lower to higher. Within the classification system of Timmers~\cite{timmers1998business}, both of our business models' primary services put the platform into the information broker model. Those focus on providing information by analyzing or finding data that can benefit the customer's operations, which is the case for both our optimization algorithms and our cloud consultancy operations. Aligning our model along the two axes Timmers' system uses, we will first find a high degree of innovation with our cloud portfolio optimization, being one of the first ever to offer this kind of service and cloud consultancy being a service that has only emerged in the past few years. Placing our business along the functional integration axis, we find that with two main functions, it falls towards the lower end of this spectrum. Both these placements fit well with the information broker classification, as seen in the~\autoref{classification_buiness_model_placement}.

\subsubsection{Classification according to Osterwalder and Pigneur}

Now we will classify our model according to the patterns of Osterwalder and Pigneur~\cite{osterwalderbusinessmodel}:

\begin{itemize}
    \item In the case of the \emph{"Unbundling"} pattern, our business model falls within the product innovation category, with a relatively new service on the market and only a few small players present in it so far. The consultancy aspect of our business model could be seen more as a customer relationship management business, meaning that, to prevent this service from conflicting with our cloud optimization service, separating them into different entities like business units may be necessary.

    \item Next, considering the so-called \emph{"Long Tail"}, we would argue that our model does not adhere to this pattern. Our business focuses on optimizing portfolios consisting of widespread cloud instances that are sold frequently, not making them niche products. We also do not offer a wide range of niche products, only a few services.

    \item On the other hand, at first glance, it can be argued that our product is a \emph{"Multi-Sided Platform"} of some kind, as it brings together two interdependent groups, and without the presence of CSPs, our platform could not exist. On the other hand, while customers optimizing their portfolio are profiting from our service, the same cannot be said for the CSPs themselves, as they rather stand to lose extra revenue generated by unused but paid-for instances running idle. Therefore, as the "Multi-Sided Platform" pattern should be of value for all involved groups, we argue that our model does not conform to this pattern.

    \item The same cannot be said for the \emph{"Free"} pattern, as our business model will include a small part of our services free of charge. While there will be various levels of subscription that will cost a monthly fee, there will also be a free trial functionality, offering limited access to the service to lure in potential customers. This approach has been labeled as "Freemium".

    \item The final pattern to take into account is the \emph{"Open"} business model. While the "inside-out" approach is not planned to be part of our business model, the "outside in" idea could potentially be explored in the future by integrating external cloud frameworks into the cloud optimization platform, which can make the product more appealing to potential customers.
\end{itemize}

\subsection{Business Model Contenders}

There are only a few business models similar to our approach.

\subsubsection{spot.io}

First, we will look at the platform spot.io\footnote{\url{https://spot.io/}}, which offers a range of tools for customers to analyze, manage and optimize their cloud portfolios. Two of their products provide functionality similar to our cloud portfolio optimization approach. Elasticgroup uses AI predictions to automate infrastructure scaling with the help of spot instances fully. On the other hand, Eco tries to optimize the customer's cloud portfolio by finding and off-loading unused Reserved Instances and Saving Plans. Furthermore, in a similar fashion to our platform, spot.io also offers consulting options to their customers, though limited to their highest subscription plan. So overall, regarding the value proposition, this firm is similar to our business model.

\subsubsection{Densify}

Next up is the platform Densify\footnote{\url{https://www.densify.com/}}, which offers a cloud management and optimization service with a similar value proposition to our platform. There is a contrast to our platform regarding revenue stream and pricing model. Densify charges the customer for each managed instance per year, with the price depending on the number of instances. Besides the high prices and seemingly not offering solutions for a portfolio of under 1000 instances, Densify does not seem to be incentivized to optimize a customer's portfolio to need fewer instances, as they charge per instance.

\subsubsection{Terraform}

The final cloud optimization platform discussed is Terraform\footnote{\url{https://www.terraform.io/}}. It allows users to express their computational infrastructure needs in their own semi-structured language, which then can be deployed to a range of resource providers like AWS or Google Cloud. The value proposition is to simplify and enable infrastructure management across multiple cloud providers. While there is also the capability of easy scalability of resource needs, there does not seem to be any optimization of the cloud portfolio. Therefore, this platform's value proposition does not directly compete with our Cloud Portfolio Manager but could work in a complementary way.

\section{Portfolio Manager Prototype}\label{cloud_portfolio_manager}

This section will present the prototype of our Cloud Portfolio Manager. It will present an overview of the various pages and showcase the various functionalities of the application.

\subsection{Login, Registration and Landing Page}
\begin{figure}[ht]
    \centering
    \includegraphics[width=0.95\linewidth]{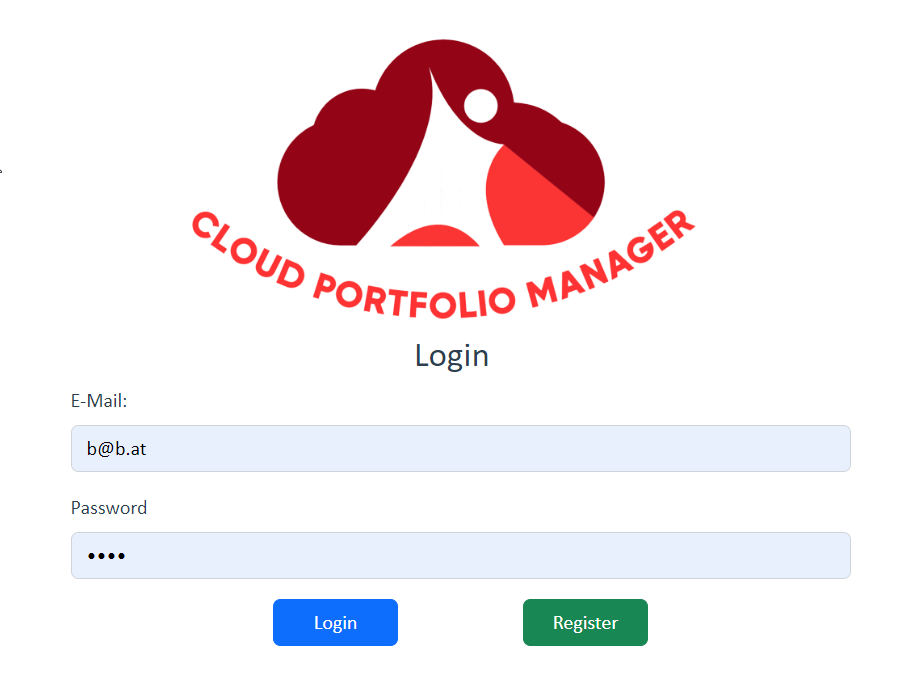}
    \caption{Login Page}
    \label{cpm_login}
\end{figure}

\begin{figure}[ht]
    \centering
    \includegraphics[width=0.95\linewidth]{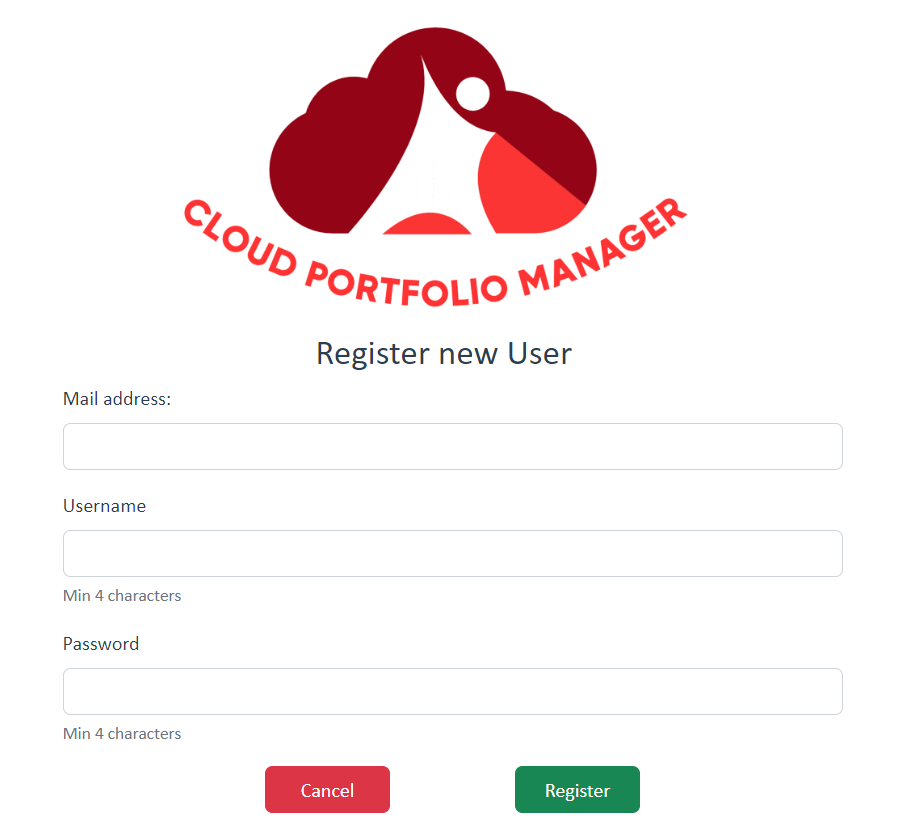}
    \caption{Registration page}
    \label{cpm_registration}
\end{figure}

Starting of, the user is presented with the login page, as seen in \autoref{cpm_login}, where the user can enter their e-mail address and password to access the website. In case a new customer does not have an account yet, the {\itshape Register} button leads to the registration page as seen in \autoref{cpm_registration}, allowing for a new account to be created by entering a valid e-mail address, a username, and a password.



Logging into the website with the correct credentials brings the customer to the landing page displaying the platform's logo and listing its creators. This page and all others besides the login and registration also feature a navbar for easy navigation between the various pages.

\subsection{Instances Page}
\begin{figure}[ht]
    \centering
    \includegraphics[width=1\linewidth]{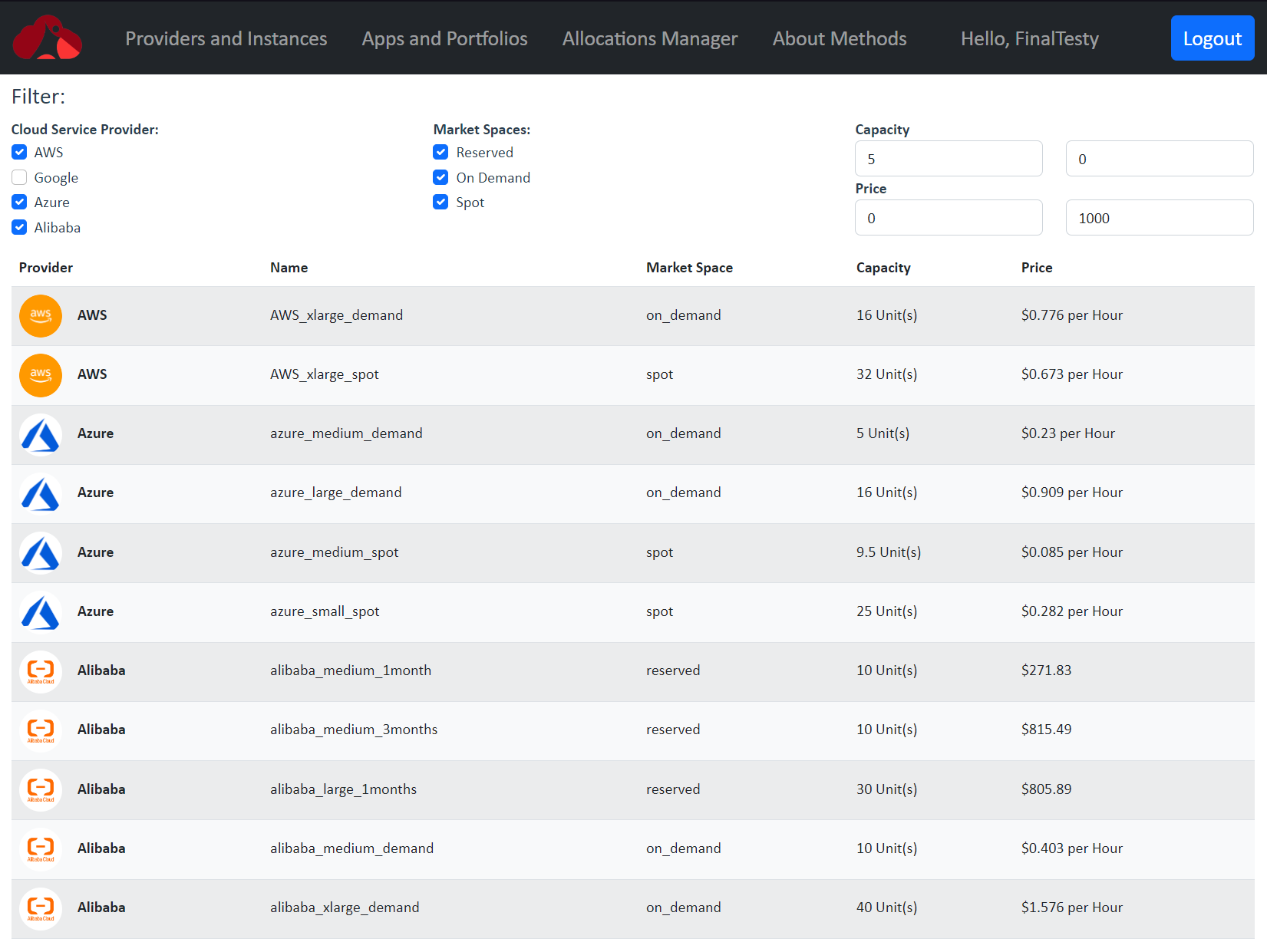}
    \caption{Instances page}
    \label{cpm_instances}
\end{figure}

This view can be navigated to through the {\itshape Providers and instances} tab in the navbar and contains information about available instances from the various providers. For our prototype, we chose a range of instances from the four biggest CSPs: AWS, Google Cloud, Microsoft Azure, and Alibaba. They all give good examples of what is available on the market. This could be adapted for future development to load instances provided by CSP APIs. The list on this page gives an overview of each instance's main attributes: provider, name, market space, capacity, and price. Furthermore, a filter function allows users to simplify their search for instances. For example, as shown in \autoref{cpm_instances}, the search does not include Google Cloud instances but instances from all market spaces, with a capacity of 5 or higher and a price of up to 1000\$.

\subsection{Apps and Portfolios Page}
\begin{figure}[ht]
    \centering
    \includegraphics[width=0.95\linewidth]{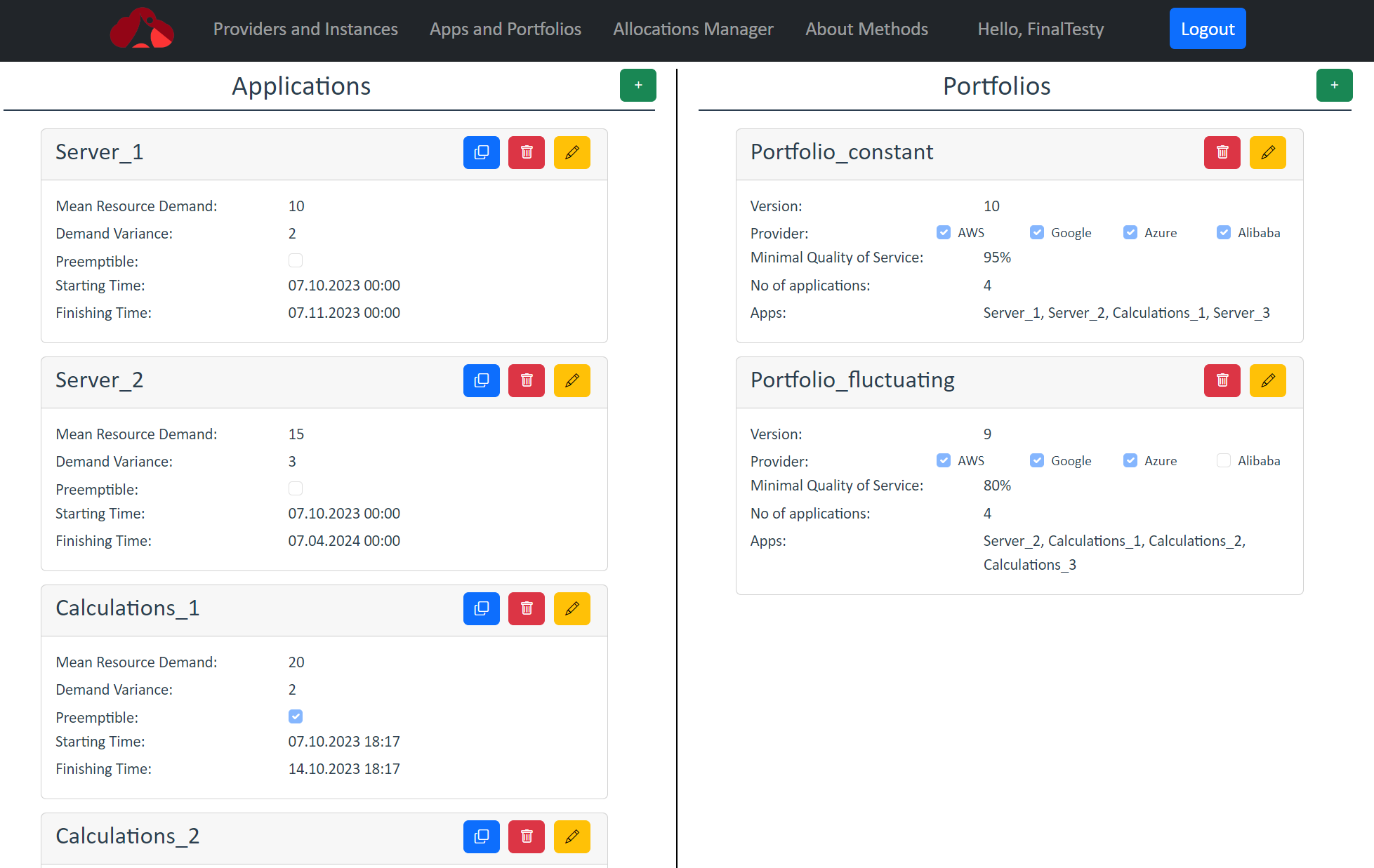}
    \caption{Apps and portfolios page overview}
    \label{cpm_apps_portfolios}
\end{figure}

Next up, the {\itshape Apps and portfolios} page enables the user to manage two of the main components of the Cloud Portfolio Platform. As seen in \autoref{cpm_apps_portfolios}, the left side lists the user's applications and details like mean resource demand, demand variance, preemtibility, and starting and finishing time. On the right side of the page is a list of the user's portfolio, including details like which providers should be considered for any possible allocation, a minimum quality of service, the number of apps in the portfolio, and a list of which applications exactly the portfolio consists of. Finally, it also states the portfolio's version, which is incremented every time a portfolio or one of its applications is changed. It is used to track which portfolio version a specific allocation has been calculated for. This enables the user to spot if any of their allocations are outdated or if any specifications or the makeup of the underlying portfolio have changed.

\begin{figure}[ht]
    \centering
    \includegraphics[width=0.95\linewidth]{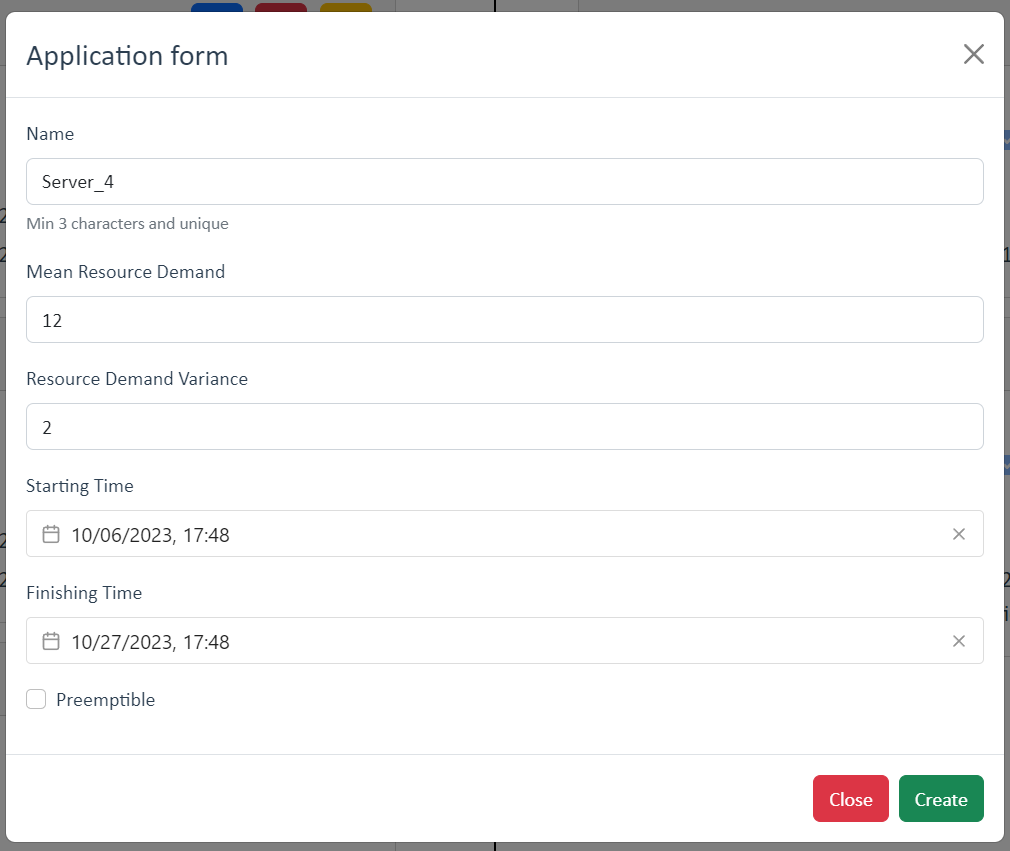}
    \caption{Application Form}
    \label{cpm_app_form}
\end{figure}

\begin{figure}[ht]
    \centering
    \includegraphics[width=0.95\linewidth]{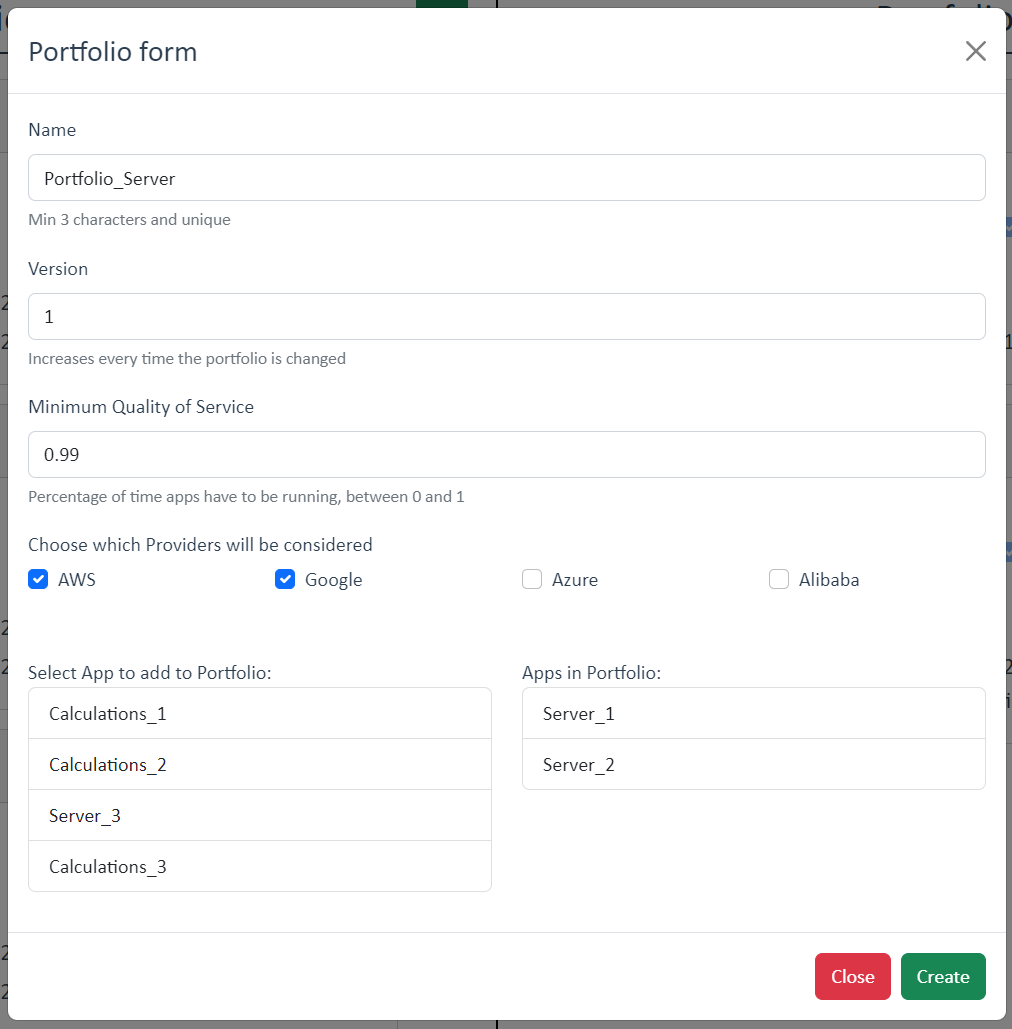}
    \caption{Portfolio Form}
    \label{cpm_portfolio_form}
\end{figure}

To create a new application or portfolio, two green buttons depict a plus sign on each side of the page. These open the respective application and portfolio forms, as seen in \autoref{cpm_app_form} and \autoref{cpm_portfolio_form}. To create an application, the user has to fill out the application form, including a unique name, mean resource demand, demand variance, a checkbox for preemtibility, and finally, the starting and finishing time chosen via a date-time picker. Should the user wish to create a portfolio, the portfolio form requires a unique name and a minimum quality of service, which gives a percentage of time the apps in the portfolio are required to run. The portfolio version, described previously, cannot be changed manually by the user. The portfolio form also requires the user to choose at least one CSP to be considered for allocations and which apps should make up the portfolio. To ensure suitable inputs for both forms, they also feature various checks, giving instant feedback to invalid inputs, such as an application's finishing time before its starting time.

The user can update each application and portfolio by clicking the yellow button, which displays a pen icon for every application and portfolio. This will open up the respective form already filled out by the app's or portfolio's data. For ease of creating several applications with similar characteristics without having to fill out the entire form every time again, applications also feature a blue copy button, which will open the application form filled out with the characteristics of the chosen application and the suffix "\_copy" added to its name. Furthermore, clicking the red button displaying a trashcan icon will delete the application or portfolio of choice. Deleting an application will also remove it from any portfolios it may be part of and increment the portfolio's version. To give feedback on operations performed on this page, creating, updating, and deleting applications and portfolios will result in a short popup denoting a successful operation or, should any errors occur, give the user notice that there has been an error. 

\subsection{Allocations Page}
This tab focuses on the primary feature of the Cloud Portfolio Manager: creating allocations for cloud portfolios. The user has a dropdown menu at the top of the page, which lists their created portfolios. Choosing a portfolio shows its details on the side, and all already existing allocations for this portfolio are below. Every allocation has an overview stating which algorithm was used, which portfolio version it was made for, its total costs, and the mean overall utilization achieved with this allocation. As allocations can take a while to be calculated, especially in the case of using the GEORG algorithm, there is also a field stating if the allocation is already completed. Below the general stats are two fields, which can be extended by clicking on them. The first contains a complete list of all instances used for this allocation and some statistics like capacity, price, and the beginning and end of the instance's run time. The second field contains more detailed statistics about the allocation, such as separate statistics for reserved, on-demand, and spot instances. Each allocation can also be deleted by clicking the red button with a trashcan icon in the header section of each entry. On the right side of the allocations list, this page also features some data visualization with graphs comparing the price, utilization overall, and utilization split by instance type for each allocation as bar charts.

\begin{figure}[ht]
    \centering
    \includegraphics[width=0.95\linewidth]{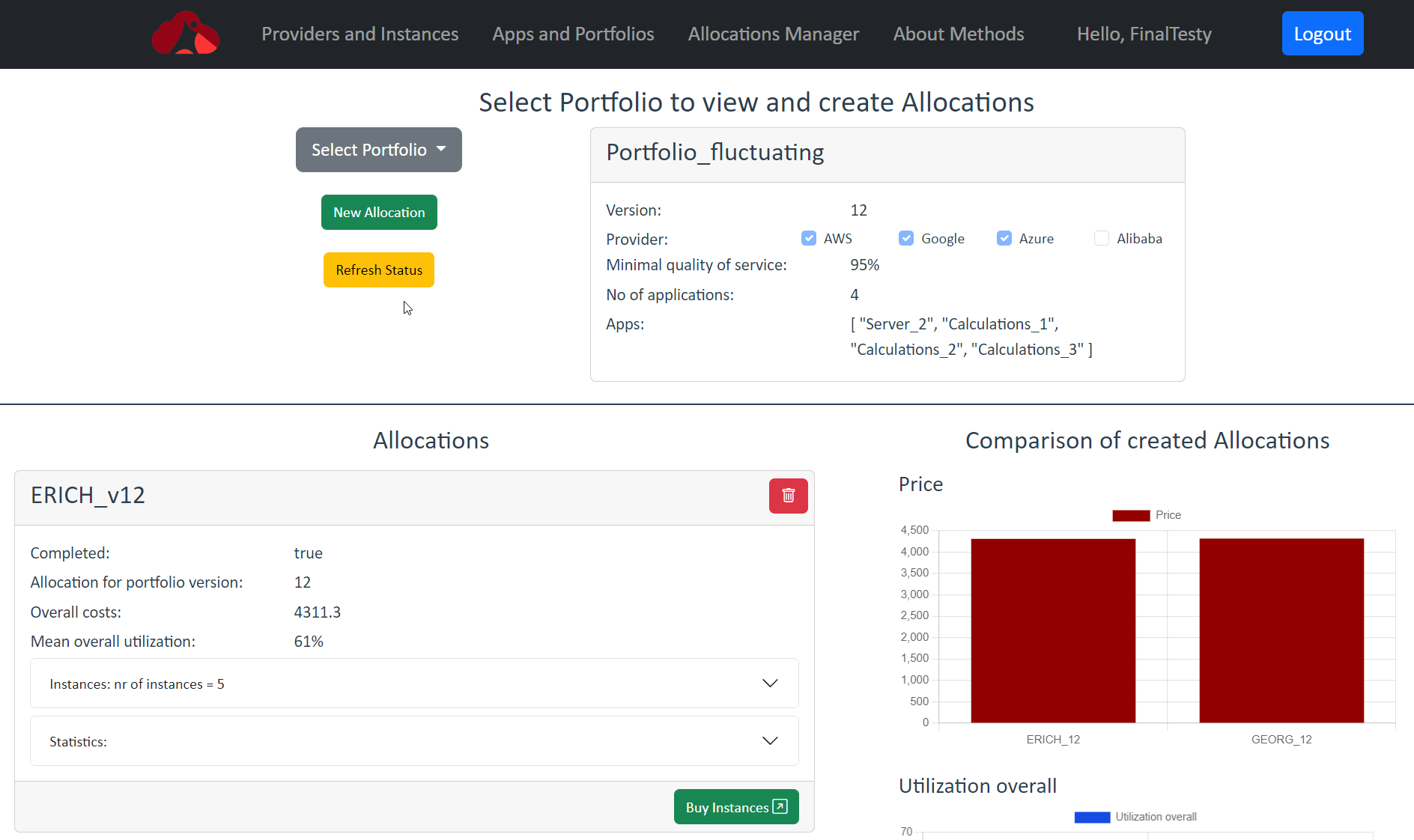}
    \caption{Allocations page overview}
    \label{cpm_allocations_overview}
\end{figure}

To create a new allocation, the "New Allocation" button opens a form where the user can choose which algorithm should be used. In the case of using ERICH, that is all that is required to do, as there is no parameterization possible. Otherwise, when selecting GEORG, various options for configuring the genetic algorithm are available. Examples of this would be the size of the population, the number of generations, and the mutation rate. Should the user want to look into this topic sparingly, there is a default value for each. Otherwise, this enables experimentation with this algorithm, which can lead to varying results. The user should be aware that this will also impact performance; for example, a very high population size is more resource-intensive for the system the platform runs on. As the calculation of the allocations is run asynchronously, there is also a "Refresh" button on this page, which reloads the allocation data from the backend.


\begin{figure}[ht]
    \centering
    \includegraphics[width=0.95\linewidth]{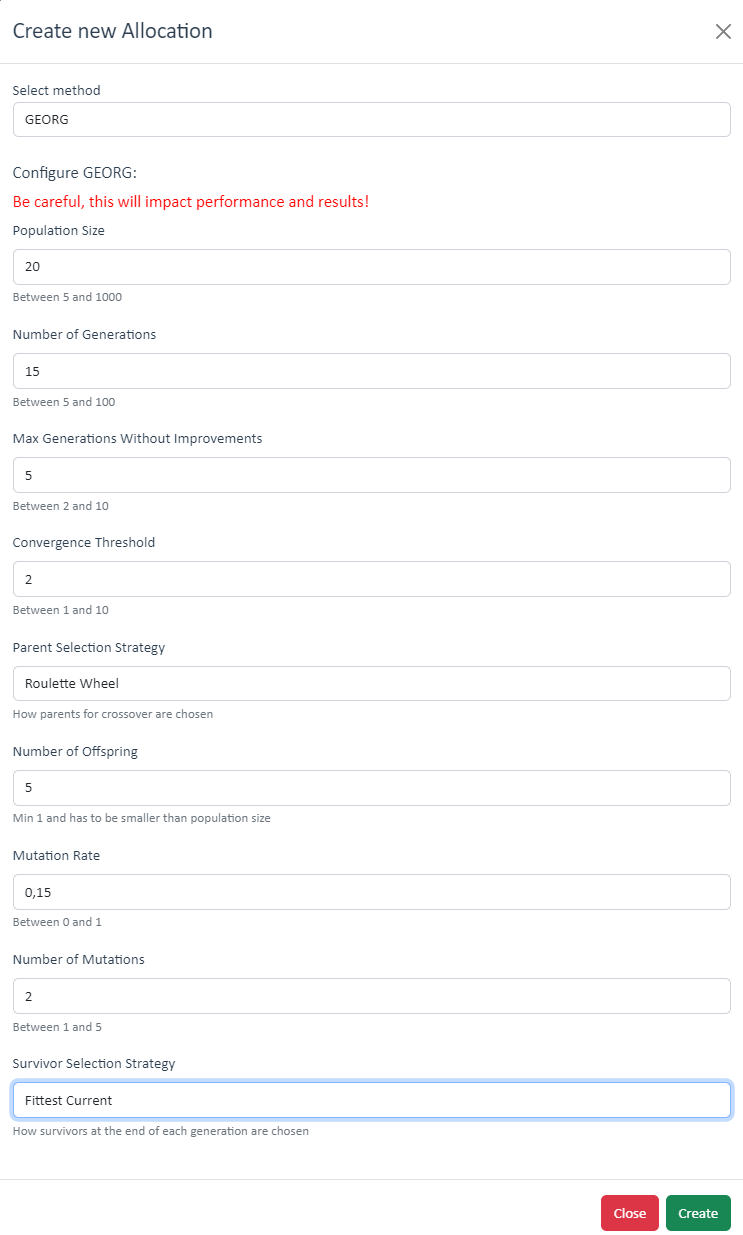}
    \caption{Form for creating a new GEORG allocation}
    \label{cpm_allocations_form}
\end{figure}

\begin{figure}[ht]
    \centering
    \includegraphics[width=0.95\linewidth]{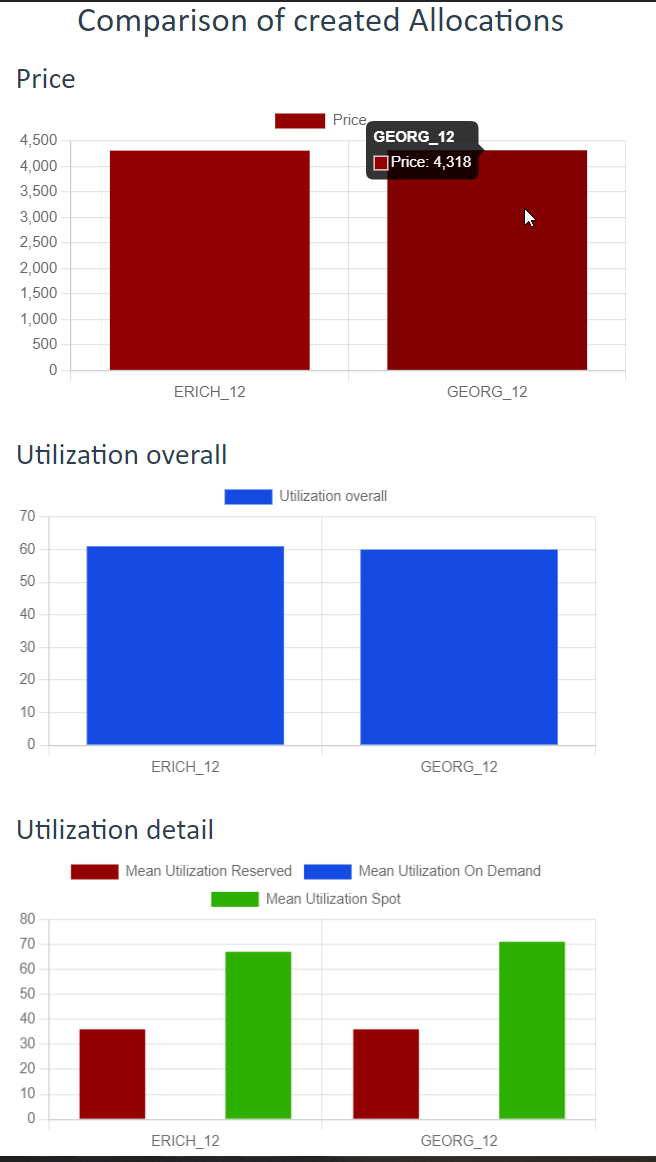}
    \caption{Graphs for comparison between a GEORG and ERICH allocation for an example portfolio}
    \label{cpm_allocations_graphs}
\end{figure}

\begin{figure}[ht]
    \centering
    \includegraphics[width=0.95\linewidth]{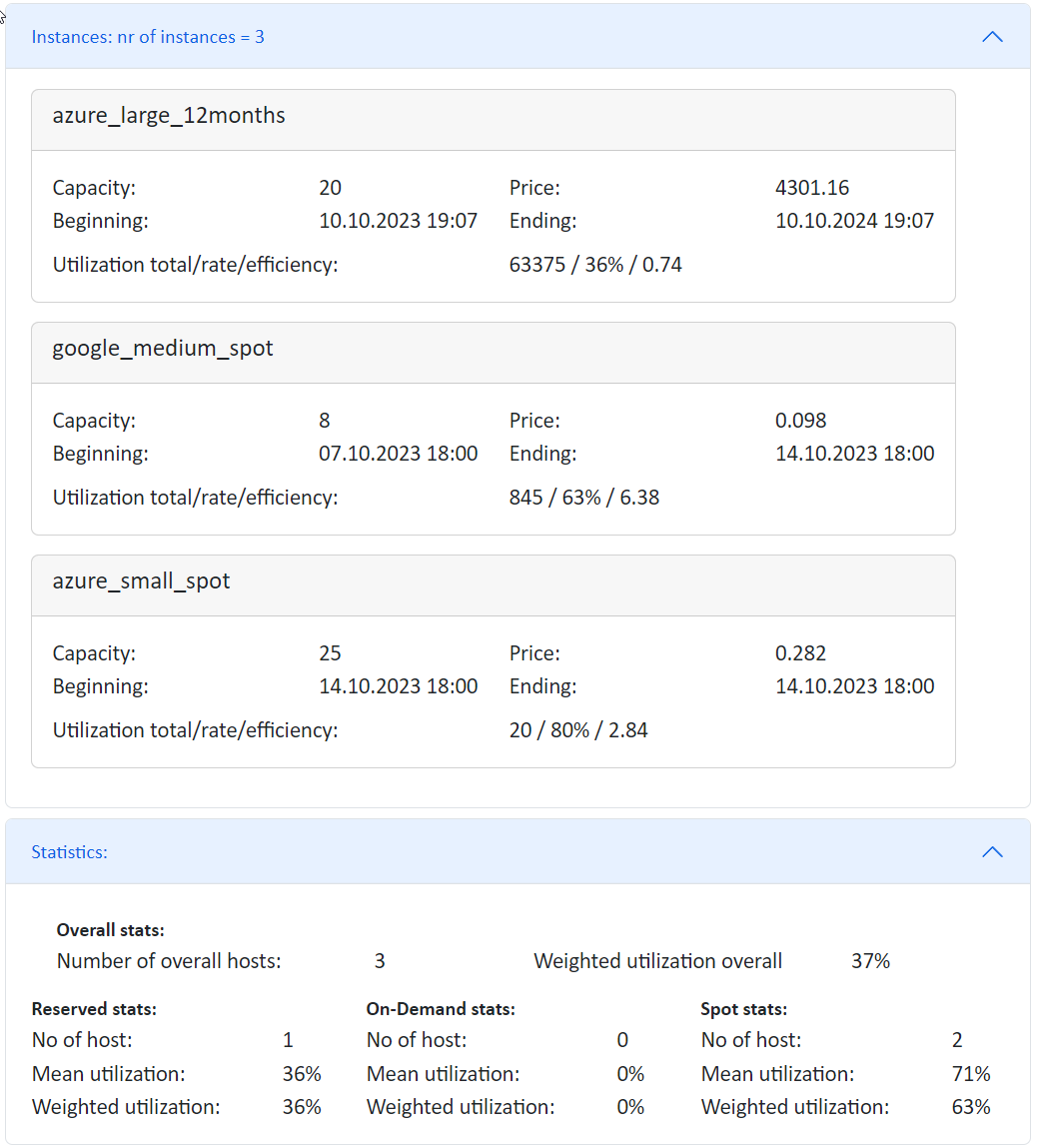}
    \caption{Details for an allocation with statistics and instances}
    \label{cpm_allocations_details}
\end{figure}

\section{User Experience Evaluation}
\label{ux_evaluation}
After the presentation of our prototype, this section will provide a brief evaluation of the user experience (UX). It is crucial for any platform targeting success in the public market to possess a convincing user interface. To this end, we will discuss the usability heuristics developed by Nielsen et al. and, based upon these same heuristics, present a brief survey on the quality of our user interface~\cite{nielsen1990heuristic,nielsen1994enhancing}.

\subsection{Usability Heuristics}
\label{usuability_heuristics}
Heuristics are an approach to finding not perfect but adequate solutions to a problem, such as the optimization approaches we developed in \autoref{cloud_portfolio_management}. This methodology can also be applied to evaluate the quality of user interfaces. One of the most commonly used and widely known usability heuristics is those developed by Nielsen in the 1990s. He proposed ten general principles for designing user interfaces, which are the following~\cite{nielsen1990heuristic,nielsen1994enhancing,nielsen_article}:

\begin{itemize}
    \item \emph{Visibility of system status}: This heuristic describes the ability of design to always communicate to the user about what is going on within the system in an easily understandable and immediate manner. A user must know what the previous interactions resulted in and understand which steps can succeed.
    
    \item \emph{Match between the system and the real world}: This refers to a design that communicates by using words and concepts that a user recognizes instead of internal terms. A design that follows this heuristic enables intuitive use of the interface without needing to learn new words or concepts.
    
    \item \emph{User control and freedom}: As users like to try various actions and mistakenly perform others, there should always be an obvious way to go back a step and cancel any action. It prevents users from getting stuck and gives them the confidence to try any action within a system freely.
    
    \item \emph{Consistency and standards}: An interface should be consistent to offer a good user experience. It refers not only to consistency within the product you offer but also to similar products a customer could be accustomed to and have developed expectations from.
    
    \item \emph{Error prevention}: This refers to designing an interface in a way that helps to prevent the occurrence of errors. Errors can be categorized into slips and mistakes. Slips are unconscious errors caused by inattentiveness and can be combated by setting helpful constraints and defaults. On the other hand, mistakes happen consciously and result from the design not properly communicating the model to the user. They can be alleviated by enabling the undoing of errors and reasonable warnings.
    
    \item \emph{Recognition rather than recall}: As users have limited short-term memory, a good design does not rely on the recall of elements, actions, and options but encourages recognition. Additional information is needed and should be easily accessible if required.
    
    \item \emph{Flexibility and efficiency of use}: This heuristic provides experienced users with possibilities to speed up interactions that inexperienced users may not need, e.g., keyboard shortcuts. These allow for flexible processes that can be executed in various ways.
    
    \item \emph{Aesthetic and minimalist design}: The design of interface elements should prioritize essential information required for their functionality. Additional irrelevant or rarely needed information competes with relevant elements for visibility.  
    
    \item \emph{Help users recognize, diagnose, and recover from errors}: If errors occur, the system should inform the user using plain language, providing an accurate description and, when possible, a suggested solution. Technical terms, such as error codes and unusual visual design for error messages, should be avoided.
    
    \item \emph{Help and documentation}: While in the best case, a system does not need any further explanation, it may be necessary to offer documentation to complete some tasks. Said documentation should be concise, easily searchable, and consist of concrete steps to be carried out.
\end{itemize}

\subsection{Methodology}
\label{ux_methodology}
We will evaluate the Cloud Portfolio Optimizer frontend design based on these heuristics. This will consist of five testers going through a set of tasks on our platform and filling out a questionnaire afterward. While the number of testers may seem relatively low, Nielsen et al. mention in their work that in contrast to one tester often missing a lot of problems, three to five aggregated evaluations offer good results~\cite{nielsen1990heuristic}.

For the evaluation process, each tester is expected to complete the following list of tasks: 
\begin{itemize}
    \item Create a new account and log into it.
    \item Navigate to the 'about methods' page and skim over the description.
    \item Go to the instances page, look at the instances displayed, and use the available filters.
    \item Navigate to the applications and portfolios page, create four new applications, update one of them, copy another, and finally delete one.
    \item Create two portfolios, update one, and delete one.
    \item Finally, proceed to the allocations page and create one allocation for each optimization approach.
    \item Log out from the platform, which returns the user to the sign-in page.
\end{itemize}

Having completed the tasks above, the testers will be asked to complete a questionnaire based on Nielsen´s heuristics. It was created using Google Forms\footnote{\url{https://docs.google.com/forms/u/0}}), and asked the user to rate the platform in regards to how well it adheres to each of the design heuristics on a scale from 1 to 5, with 5 representing the best possible adherence. Furthermore, the testers were questioned on whether they found any issues or had suggestions regarding each heuristic. To gauge the testers' expertise in regards to IT in general and the topic of cloud markets, the survey also contains a self-assessment of these topics. 

\subsection{Results}
\label{ux_results}
The questionnaire results gave helpful feedback on the design of the user interface and pointed out a couple of errors that were overlooked in development. Five testers completed the survey as described in \autoref{ux_methodology}. Regarding IT-related experience, two testers reported no background in IT, one mentioned having educational knowledge, and two stated having work-related IT experience. The testers' knowledge about the cloud market was somewhat limited, with three testers stating their understanding to be cursory and two having no knowledge about the topic.

Overall, the platform was perceived as adhering to most heuristics pretty well, with some achieving better results than others. The best-rated heuristics were those of "aesthetic and minimalist design" as well as "user control and freedom," achieving an average of 4.6 and 4.4 points, respectively. The heuristics of "consistency and standards" and "recognition and standards" also scored well, averaging 4.0 points. All other heuristics got an average of 3.4 or 3.8, with the notable exception of "error prevention" only scoring 3.0 on average. The complete list of average scoring can be seen in \autoref{questionnaire_statistics}.

\begin{table}[ht]
\begin{center}
    \begin{tabular}{ | l | l | }
    \hline
    Visibility of system status & 3.8 \\ \hline
    Match between the system and the real world & 3.4 \\ \hline
    User control and freedom & 4.4 \\ \hline
    Consistency and standards & 4.0 \\ \hline
    Error prevention & 3.0 \\ \hline
    Recognition rather than recall & 4.0 \\ \hline
    Flexibility and efficiency of use & 3.8 \\ \hline
    Aesthetic and minimalist design & 4.6 \\ \hline
    Help users recognize, diagnose, and recover from errors & 3.4 \\ \hline
    Help and documentation & 3.4 \\ \hline
    \end{tabular}
\caption{Table of average assessment for each heuristic}
\label{questionnaire_statistics}
\end{center}
\end{table}

These results point towards the platform having an aesthetic and mostly easy-to-use design, with room for improvement in error prevention and handling, as well as documentation and help. All testers, except for one, who provided no useful feedback by rating several heuristics poorly without offering any comments, also reported issues they encountered and provided suggestions for improvement. 
%

To sum up, the testers rated the platform positively regarding most heuristics and provided helpful feedback for improvement. The suggested feedback concerning minor changes has already been applied, while others inform future ways of improving the platform, such as adding a user guide.

\section{Conclusion and Future Work}\label{conclusion_future_work}

This paper focuses on cloud portfolio management platforms and related business models. Cloud portfolio management is primarily concerned with finding cost-efficient allocations for cloud resources. Having discussed the necessary concepts around the work built, we proposed our business model for a Cloud Portfolio Manager. With the business model in place, we implemented a prototype of the Cloud portfolio manager~\cite{haagmaster24}, incorporating two optimization algorithms that had been developed~\cite{CPM_Optimization}.

As the cloud computing market has been on a meteoric rise over the past years and is still expanding, the topic of cloud portfolio management will likely keep or expand its relevance in the coming years. Future work would see the prototype developed into a fully functional public platform operating with our business model or a modification of it. Furthermore, the platform's offered services could be extended into various monitoring functionalities and direct control of portfolios via the platform. Further work will improve the existing optimization algorithms and create new ones. In addition to tighter packing, these could, for example, elaborate on the resource constraints considered, such as network capabilities and storage. Finally, further research will build upon the business model presented in this thesis. Thus, we work on integrating the presented approach into our work on automatic and dynamic resource (re-)negotiation and contracting of resources between providers and customers~\cite{schikuta_mach_negotiation}.



\end{document}